\documentclass[useAMS,usenatbib]{mn2e}

\usepackage{graphicx} 
\usepackage{amssymb}
\usepackage{journals}
\usepackage{placeins}
\usepackage[modulo]{lineno}
\usepackage{hyperref}
\usepackage{rotating}
\usepackage{CJK}
\newcommand{\teff}{$T_{\rm eff}$} 
\newcommand{\tcond}{$T_{\rm cond}$} 
\newcommand{\logg}{$\log g$} 

\newcommand{\vt}{$\xi_t$}

\newcommand\mjup{M$_{\rm Jup}$}

\title[Detailed elemental abundances of binary stars]{Detailed elemental abundances of binary stars: Searching for signatures of planet formation and atomic diffusion\thanks{Based on observations collected at the European Organisation for Astronomical Research in the Southern Hemisphere under ESO Programme 0101.D-0921(A). Part of the data presented herein were obtained at the W.M.\ Keck Observatory, which is operated as a scientific partnership among the California Institute of Technology, the University of California and the National Aeronautics and Space Administration. The Observatory was made possible by the generous financial support of the W.M.\ Keck Foundation.}}
\author[F.\ Liu et al.]
{Fan Liu (刘凡),$^{1}$\thanks{E-mail: fanliu@swin.edu.au}
Bertram Bitsch,$^{2}$
Martin Asplund, 
Bei-Bei Liu (刘倍贝),$^{3,4}$ \newauthor
Michael T.\ Murphy,$^{1}$
David Yong,$^{5}$
Yuan-Sen Ting (丁源森)$^{5,6,7,8}$ and
Sofia Feltzing$^{3}$ \\
\\
$^{1}$Centre for Astrophysics and Supercomputing, Swinburne University of Technology, Melbourne, VIC 3122, Australia\\
$^{2}$Max-Plank-Institute f\"ur Astronomy, K\"onigstuhl 17, 69117 Heidelberg, Germany\\
$^{3}$Lund Observatory, Department of Astronomy and Theoretical physics, Lund University, Box 43, SE-22100 Lund, Sweden\\
$^{4}$Department of Physics, Zhejiang University, Hangzhou 310027, China\\
$^{5}$Research School of Astronomy and Astrophysics, Australian National University, Canberra, ACT 2611, Australia\\
$^{6}$Institute for Advanced Study, Princeton, NJ 08540, USA\\
$^{7}$Department of Astrophysical Sciences, Princeton University, Princeton, NJ 08540, USA\\
$^{8}$Observatories of the Carnegie Institution of Washington, Pasadena, CA 91101, USA
}

\begin{document}
\begin{CJK*}{UTF8}{gbsn}

\date{Accepted ? Received ?; in original form 2020 December 28}

\pagerange{\pageref{firstpage}$-$\pageref{lastpage}} \pubyear{2021}

\maketitle

\label{firstpage}

\begin{abstract}
Binary star systems are assumed to be co-natal and coeval, thus to have identical chemical composition. In this work we aim to test the hypothesis that there is a connection between observed element abundance patterns and the formation of planets using binary stars. Moreover, we also want to test how atomic diffusion might influence the observed abundance patterns.
We conduct a strictly line-by-line differential chemical abundance analysis of 7 binary systems. Stellar atmospheric parameters and elemental abundances are obtained with extremely high precision ($<$\,3.5\%) using the high quality spectra from VLT/UVES and Keck/HIRES. We find that 4 of 7 binary systems show subtle abundance differences (0.01 - 0.03 dex) without clear correlations with the condensation temperature, including two planet-hosting pairs. The other 3 binary systems exhibit similar degree of abundance differences correlating with the condensation temperature.
We do not find any clear relation between the abundance differences and the occurrence of known planets in our systems. Instead, the overall abundance offsets observed in the binary systems (4 of 7) could be due to the effects of atomic diffusion. Although giant planet formation does not necessarily imprint chemical signatures onto the host star, the differences in the observed abundance trends with condensation temperature, on the other hand, are likely associated with diverse histories of planet formation (e.g., formation location). Furthermore, we find a weak correlation between abundance differences and binary separation, which may provide a new constraint on the formation of binary systems.
\end{abstract}

\begin{keywords}
stars: abundances -– stars: atmospheres -– stars: evolution -- stars: binaries: general -- stars: planetary systems -- planets and satellites: formation
\end{keywords}

\section{Introduction}

Thousands of exoplanets around solar-type stars have been detected via radial velocity and transit surveys during the last two decades. The properties of these planetary systems (i.e. mass, radius, distance from host star) exhibit a surprisingly enormous degree of diversity \citep{wf15}. The diversity of exoplanets, in another sense, may correspond to different variations during the formation processes \citep{bit19,izi19,lam19,liubb19,lj20}, making it complicated to examine the real influence of planet formation on the host star's chemical composition \citep{liu20}. In addition to the well-established correlation between stellar metallicity and giant planet occurrence rate \citep{gon97,san04,fv05}, a few hypotheses were proposed to build further connections between the elemental abundances of planet-hosting stars and the process of planet formation but none of them are conclusive.

In particular, \citet[hereafter M09]{mel09} proposed that the formation of terrestrial planets can take away more refractory material (elements with high condensation temperature \tcond) than volatile material (low \tcond\ elements) from the proto-planetary disc during the early accretion stage. This process thus results in lower abundances in refractory elements than volatile elements of a star hosting rocky planets. This hypothesis is favoured in explaining the peculiar abundance pattern of the Sun - deficiency in refractory elements, when compared to the majority of solar twins (M09, \citealp{ram10,bed18}). Following M09 scenario, several studies discussed in theory the impact of different properties and formation processes of exoplanets (e.g., \citealp{cha10,bit18,kun18}). Among these studies, \citet{bit18} demonstrated that different formation location and history of a super-Earth or ice-giants (inside/outside H$_2$O or CO ice-line) should play a key role in determining the observed abundance pattern of planet-hosting systems, which can be tested by comparing binary stars with or without planets.

A recent study by \citet{bo20}, however, attributed the observed abundance pattern of the Sun to the formation of Jupiter analogs (i.e. giant planets at large separations). In their scenario, forming a giant planet can create a gap that traps a large amount of exterior dust in early stage, preventing them being accreted by the host star in contrast to the gas, thus leading up to a 10\% - 15\% refractory depletion\footnote{The exact depletion would depend on the formation time of the giant planet and when it starts to trap the dust exterior to its orbit.}. 

A tentative explanation for the abundance signatures found in M09 was proposed by \citet{gus18a,gus18b}. They suggested that the observed solar chemical composition may be due to the dust cleansing of the proto-solar cloud. Their scenario associates the peculiar solar abundance patterns to its birth environment. 

Alternatively, post-formation accretion of inner planetary material, namely planet engulfment, can also pollute the stellar chemical composition \citep{cha01, pin01}. For a stellar-planetary system with the history of strong dynamical interaction, it is possible that the inner planets or material have fallen into the host star, thus enhancing the stellar surface abundances (see e.g., \citealp{chu20}). A \tcond-dependent trend, where most refractory elements being over-abundant relative to the volatile elements, can be induced by this process. However, \citet{tv12} demonstrated that this is unlikely due to the rapid erasure of such metal-rich material from the convection zone, as a result of thermohaline mixing. Although on theoretical grounds chemical signatures of planet engulfment are unexpected, various observational studies have suggested it as a possibility (e.g., HAT-P-4\,A/B from \citealp{saf17}, and HD\,240429/30 from \citealp{oh18}).

In most formation channels, binary star systems with wide separations (a hundred AU to 1\,pc) are assumed to form from the same gas at about the same time \citep{off10,rm12,kou10,tok17}, and share identical chemical composition. Therefore, any relative abundance differences between binary components likely reflect astrophysical processes including signatures of planetary formation, and provide clues to test the hypotheses of planet formation. High-precision spectroscopic studies have been applied to two dozens of solar-type binary systems (with/without known planets). These studies have attempted to measure relative elemental abundances in each binary system, where both components are similar to each other in terms of their atmospheric parameters. This approach minimises the systematic uncertainties in a strictly differential abundance analysis, enabling the investigation of subtle abundance anomalies.

The correlation between differences in elemental abundances of two binary components and \tcond\ was reported in \citet{tuc14,tuc19} and \citet{ram15} for the 16\,Cygni\,A/B and XO-2 systems. A similar abundance pattern was detected in another binary system (HD 20807/20766), in which one of the components is surrounded by a debris disc \citep{saf16}. These studies favour the M09 scenario, where they found the planet-hosting stars are depleted in refractory elements. Meanwhile, \citet{saf17,oh18,ram19} and \citet{nag20} have studied several binary systems and found that one of the binary components exhibits a significant increase in elemental abundances (0.1 - 0.2 dex) of e.g., HAT-P-4 system, HD\,240430/29 and HIP\,34407/26. They tentatively attributed the large abundance differences to the effects of planet engulfment. Furthermore, subtle or moderate abundance differences ($\lesssim$\,0.03 dex) without clear $T_{\rm cond}$ trend were reported in several binary systems (e.g., WASP-94\,A/B, HD\,133131A/B, HD\,106515A/B, HD\,20781/20782, HAT-P-1 system and HD\,80606/07) by \citet{tes16a}, \citet[hereafter T16]{tes16b}, \citet[hereafter S19]{saf19}, \citet{mac14}, \citet{liu14} and \citet{liu18}, respectively. These studies may demonstrate that hosting close-in giant planets does not necessarily alter the stellar surface abundances, although no conclusive interpretation was offered. To summarise, it remains unclear how and when planet formation leaves an imprint in the observed elemental abundance differences between binary components.

Unveiling the subtle abundance differences within binary systems not only allows us to examine the hypotheses about planet formation, but also helps us to test the chemical homogeneity of binary systems as 'mini-clusters'. Recent studies have attempted to characterise and understand the subtle elemental abundance variations of systems that expected to have common composition, such as binaries \citep{and19,haw20,nel21} and open clusters \citep{liu16,spi18}. The deviation from chemical homogeneity might indicate that the interstellar medium (ISM) mixing is not homogeneous. In fact, the subtle correlation between the deviation among elemental abundances has been proposed to be an useful probe to study both stellar nucleosynthesis and ISM mixing \citep{tw21}.

In this paper we study in detail the elemental abundances of 7 binary star systems with high-precision differential analysis and search for signatures of planet formation and atomic diffusion. In Section 2 we describe the sample selection, observations and data reduction. The differential stellar atmospheric parameters and chemical abundances of our programme binaries are presented in Section 3. We discuss the potential hypotheses of planet formation and the atomic diffusion effects in Section 4, and summarise the conclusions in Section 5.

\section{Sample selection, observations and data reduction}

Our targets were initially selected from \citet{des04,des06}, with observability in the proposed semester being considered. We selected binary systems with projected separation $>$\,300\,AU ($>$\,5") so that they are spectroscopically distinct and the proto-planetary discs around the individual stars, if present, can not exchange mass. We required that both components of each pair in our sample have magnitude of Gaia G $<$ 9.2\,mag, colour of 0.6 $<$ Gaia (BP$-$RP) $<$ 1.0 and colour difference of $\Delta$(BP$-$RP) $<$ 0.1. This would facilitate high-precision differential abundance analysis since the binary components are similar to each other.

\begin{figure}
\centering
\includegraphics[width=\columnwidth]{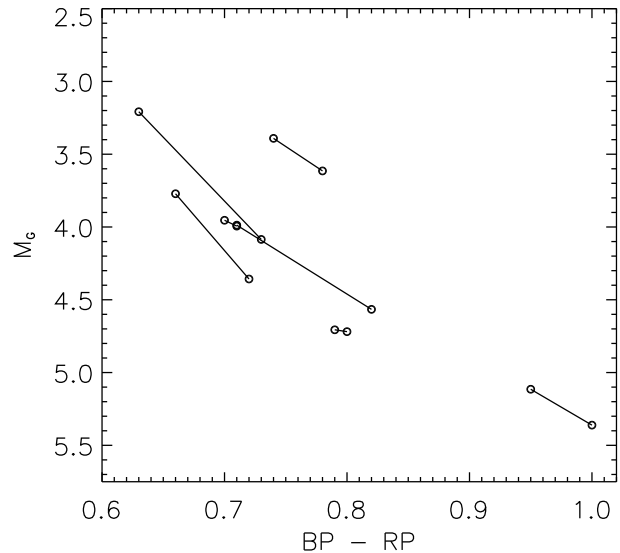}
\caption{Colour-magnitude diagram of our programme binary systems: Gaia (BP$-$RP) as a function of absolute magnitude M$_G$. The two components of each pair are linked by solid lines.}
\label{fig:cmd}
\end{figure}

In total we obtained high resolution, high $S/N$ spectra of 7 binary systems (14 stars), of which two pairs host known giant planet(s) (HD\,106515A/B and HD\,133131A/B). The observational information of our objects are listed in Table \ref{t:info}, where the photometric and astrometric data were taken from Gaia Data Release 2 \citep{gai18}. The 'A' and 'B' components of each pair are labelled in the table for clarification. The colour-magnitude diagram of our objects is shown in Figure \ref{fig:cmd}. We note the differences in both colours and absolute magnitudes are $\lesssim$\,0.1\,mag between binary components in our sample.

\begin{table*}
\caption{Observational information of the programme binary systems. The photometric and astrometric data were adopted from Gaia Data Release 2 \citep{gai18}.}
\centering
\label{t:info}
\begin{tabular}{@{}lrrcccrrccc@{}}
\hline\hline
Object & RA & Dec & Gaia\,G & BP$-$RP & Parallax & PMRA$^a$ & PMDEC$^a$ & M$_{\rm G}$ & Instrument & Planets \\
 & (deg) & (deg) & (mag) & (mag) & (mas) & (mas/yr) & (mas/yr) & (mag) & & \\
\hline
HD\,106515A & 183.7763 & $-$7.2575 & 7.78 & 0.95 & 29.304 & $-$251.58 & $-$51.39 & 5.11 & VLT/UVES & 1$^b$ \\
HD\,106515B & 183.7744 & $-$7.2577 & 8.02 & 1.00 & 29.392 & $-$244.68 & $-$67.93 & 5.36 & VLT/UVES & ... \\
HD\,108574$^c$ & 187.0174 & 44.7943 & 7.28 & 0.71 & 21.951 & $-$182.13 & $-$4.69 & 3.99 & Keck/HIRES & ... \\
HD\,108575$^d$ & 187.0189 & 44.7918 & 7.86 & 0.82 & 21.937 & $-$180.39 & 0.44 & 4.57 & Keck/HIRES & ... \\
HD\,111484A & 192.3947 & 3.8182 & 8.66 & 0.70 & 11.449 & $-$79.88 & $-$4.47 & 3.95 & VLT/UVES & ... \\
HD\,111484B & 192.3955 & 3.8206 & 8.70 & 0.71 & 11.450 & $-$76.70 & $-$3.73 & 3.99 & VLT/UVES & ... \\
HD\,133131A & 225.8985 & $-$27.8431 & 8.28 & 0.80 & 19.399 & 155.99 & $-$133.72 & 4.72 & VLT/UVES & 2$^b$ \\
HD\,133131B & 225.9000 & $-$27.8416 & 8.26 & 0.79 & 19.465 & 158.95 & $-$139.10 & 4.71 & VLT/UVES & 1$^b$ \\
HD\,146368A & 244.3625 & $-$34.8189 & 8.39 & 0.66 & 11.921 & 40.07 & 81.28 & 3.77 & VLT/UVES & ... \\
HD\,146368B & 244.3610 & $-$34.8182 & 9.00 & 0.72 & 11.788 & 34.04 & 83.99 & 4.36 & VLT/UVES & ... \\
HD\,216122A & 342.4226 & 40.5151 & 7.96 & 0.74 & 12.196 & 1.08 & $-$37.00 & 3.39 & Keck/HIRES & ... \\
HD\,216122B & 342.4206 & 40.5148 & 8.06 & 0.78 & 12.911 & $-$3.63 & $-$26.46 & 3.61 & Keck/HIRES & ... \\
HIP\,92961$^c$ & 284.0887 & 45.5150 & 9.15 & 0.73 & 9.707 & 22.20 & 59.08 & 4.09 & Keck/HIRES & ... \\
HIP\,92962$^d$ & 284.0923 & 45.5073 & 8.26 & 0.63 & 9.764 & 24.99 & 58.52 & 3.21 & Keck/HIRES & ... \\
\hline
\end{tabular}
\\
$^a$ PMRA and PMDEC: proper motion in Right Ascension and Declination.\\
$^b$ Giant planets detected by radial velocity method.\\
$^c$ The 'A' component of binary. \enspace \enspace \enspace
$^d$ The 'B' component of binary.\\
\end{table*}

Three programme binary systems (HD\,108574/75, HD\,216122A/B and HIP\,92961/62) were observed on 2016-05-18, using the High Resolution Echelle Spectrometer (HIRES; \citealt{vog94}) mounted on the Keck\,I telescope\footnote{Keck Programme ID: Z007Hr, 2016A.}. We chose the E4 slit (width of 0\farcs4) which corresponds to a spectral resolving power of $R \approx$ 86,000. The obtained spectra cover a range of wavelength from 400\,nm to 800\,nm. The MAuna Kea Echelle Extraction ({\sc MAKEE}) data reduction pipeline was applied to reduce the raw Keck/HIRES spectra.

Four programme binary systems (HD\,106515A/B, HD\,111484A/B, HD\,133131A/B and HD\,146368A/B) were observed using the ESO UV-visual echelle spectrograph (UVES; \citealt{dek00}), assembled on the ESO's Very Large Telescope (VLT-UT2)\footnote{ESO Programme ID: 0101.D-0921(A).}. We used the standard \textit{DIC-2 437$+$760} setting with the image slicer\#3. The obtained UVES spectra have a spectral resolving power of $R \approx$ 100,000, and cover a wavelength range of 380\,nm - 500\,nm and 600\,nm - 800\,nm. We made use of the {\sc EsoReflex} pipeline \citep{fre13} to reduce the raw data.

\begin{figure}
\centering
\includegraphics[width=\columnwidth]{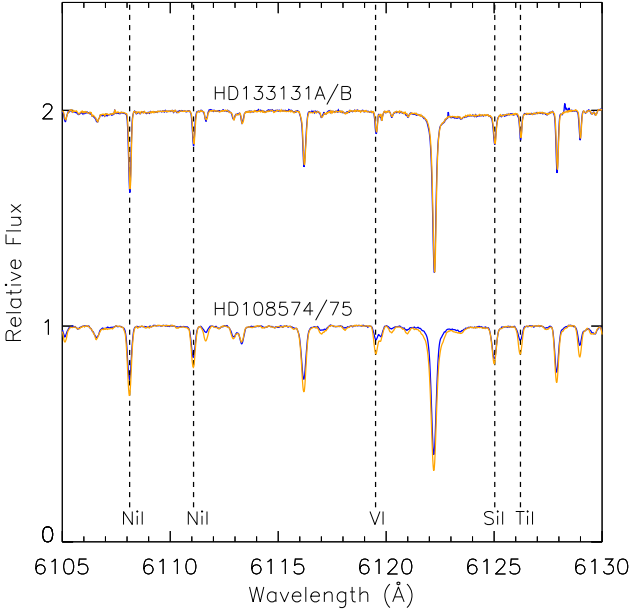}
\caption{A portion of the normalised spectra of two programme binary systems: HD\,133131A/B and HD\,108574/75. Blue: spectra of HD\,133131A and HD\,108574; orange: spectra of HD\,133131B and HD\,108575. The dashed lines mark out several atomic lines (Si\,{\sc i}, Ti\,{\sc i}, V\,{\sc i} and Ni\,{\sc i}) adopted in our analysis.}
\label{fig:spec}
\end{figure}

The reduced individual spectra were radial-velocity corrected, co-added and normalised using {\sc IRAF}\footnote{{\sc IRAF} is distributed by the National Optical Astronomy Observatory, which is operated by the Association of Universities for Research in Astronomy, Inc., under cooperative agreement with the National Science Foundation.}. The combined spectrum of each binary component attained extremely high $S/N$ ($>$\,400 per pixel near 600\,nm). A portion of the reduced spectra of two binary systems (HD\,133131A/B and HD\,108574/75) is shown in Figure \ref{fig:spec} as an example. The spectral quality is suggested to be proportional to R $\times$ $(S/N)$ at a given wavelength \citep{nor01}. Therefore the quality of our spectra are at least 2 - 3 times better than the previous available data (e.g., HD\,106515A/B and HD\,133131A/B, both with R $\lesssim$ 65,000 and S/N $\lesssim$ 250). Only with such high-quality spectra are we able to achieve the abundance precision of 0.01 dex, which allow us to detect and disentangle the subtle abundance differences due to e.g., planet formation and atomic diffusion.

The line-lists from \citet{mel14,liu14,liu18} were combined to provide a master list of more than two hundred non-blended spectral lines of up to 30 elements (C, O, Na, Mg, Al, Si, S, K, Ca, Sc, Ti, V, Cr, Mn, Fe, Co, Ni, Cu, Zn, Sr, Y, Zr, Ba, La, Ce, Nd, Sm, Eu, Gd and Dy). 
When analysing a single binary, we selected the lines from the master list that could be used to analyse the spectra of both stars; this list of lines used is slightly different for different binaries. The equivalent widths (EWs) of selected spectral lines were measured using the {\sc IRAF/splot} function manually with Gaussian or Lorentz profile fitting. For a single pair, we set the local continuum to be consistent for each spectral line of both components. The method is similar to \citet{liu16,liu18}, which leads to precise measurements on a differential basis. In order to avoid the adverse effects of line saturation, we excluded strong lines with EW $\geq$ 120\,m\AA\, except for a few inevitable elements (Mg\,{\sc i}, Mn\,{\sc i,} and Ba\,{\sc i}). The EW measurements and atomic data of adopted spectral lines for this analysis are presented in Table A1. We note that the influence of adopted atomic data (e.g., $gf$-values) on a line-by-line differential abundance analysis of similar stars are negligible (e.g., \citealp{liu14,liu18}).

\section{Analysis and results}

\subsection{Differential stellar atmospheric parameters}

We addressed an elemental abundance analysis using the {\sc MOOG\,2014} software \citep{sne73,sob11} on the basis of 1D local thermodynamic equilibrium (LTE). The model atmospheres were derived from the Kurucz {\sc ODFNEW} grid \citep{ck03}. For each binary system, we obtained the stellar atmospheric parameters (i.e. effective temperature \teff, surface gravity \logg, microturbulent velocity \vt\ and metallicity [Fe/H]) of both components with the traditional method, where an excitation and ionization balance were met using the abundances of Fe\,{\sc i} and Fe\,{\sc ii} lines.

We adopted the derived stellar atmospheric parameters of the 'A' component as the reference for each pair. The differential stellar parameters of the 'B' component relative to the 'A' component were then established separately using an iterative grid-search algorithm described in \citet{liu14}. Briefly speaking, a best combination of differential stellar parameters from each iteration was obtained from a grid of stellar atmospheric models which is successively refined. The solution minimises the difference between $\Delta$[Fe\,{\sc i}/H] and $\Delta$[Fe\,{\sc ii}/H], as well as the trend of $\Delta$[Fe\,{\sc i}/H] versus reduced EW ($\log$\,(EW/$\lambda$) and lower excitation potential (LEP). We also required the adopted metallicity from the stellar atmospheric model to be consistent with the calculated average [Fe/H]. We decreased the grid's step-size into half after each iteration, and determined the final solution when the step-size was down to \teff\ = 1\,K, \logg\ = 0.01 dex\footnote{Unit in log(cm\,s$^{-2}$).}, and \vt\ = 0.01\,km\,s$^{-1}$. We clipped the iron lines with differential abundances out of $> 3\,\sigma$ in the final run. The adopted stellar parameters, on a differential basis, fulfil the balance of excitation and ionization. Table \ref{t:para} lists the final stellar atmospheric parameters of each pair in our sample. These binary systems have differences up to \teff\ of $\pm$ 400\,K, \logg\ of $\pm$ 0.2 dex, and [Fe/H] of only $\pm$ 0.03 dex.

Following \citet{eps10} and \citet{ben14}, we derived the uncertainties in the differential stellar parameters by taking into account the co-variances between changes in the stellar atmospheric parameters and the differential abundances of Fe\,{\sc i} and Fe\,{\sc ii}. We achieved very high precision because of the use of high-quality spectra and the application of strictly line-by-line differential method, which reduces remarkably the systematic errors from deficiencies in the modelling (1D LTE) of the stellar atmospheres, formation of spectral lines and atomic line data (see e.g., \citealp{jof19} and \citealp{asp21}). The typical uncertainties in our differential stellar parameters are: $\sim$ 15\,K for \teff, $\sim$ 0.04 dex for \logg, and $\sim$ 0.014 dex for [Fe/H].

\begin{figure}
\centering
\includegraphics[width=\columnwidth]{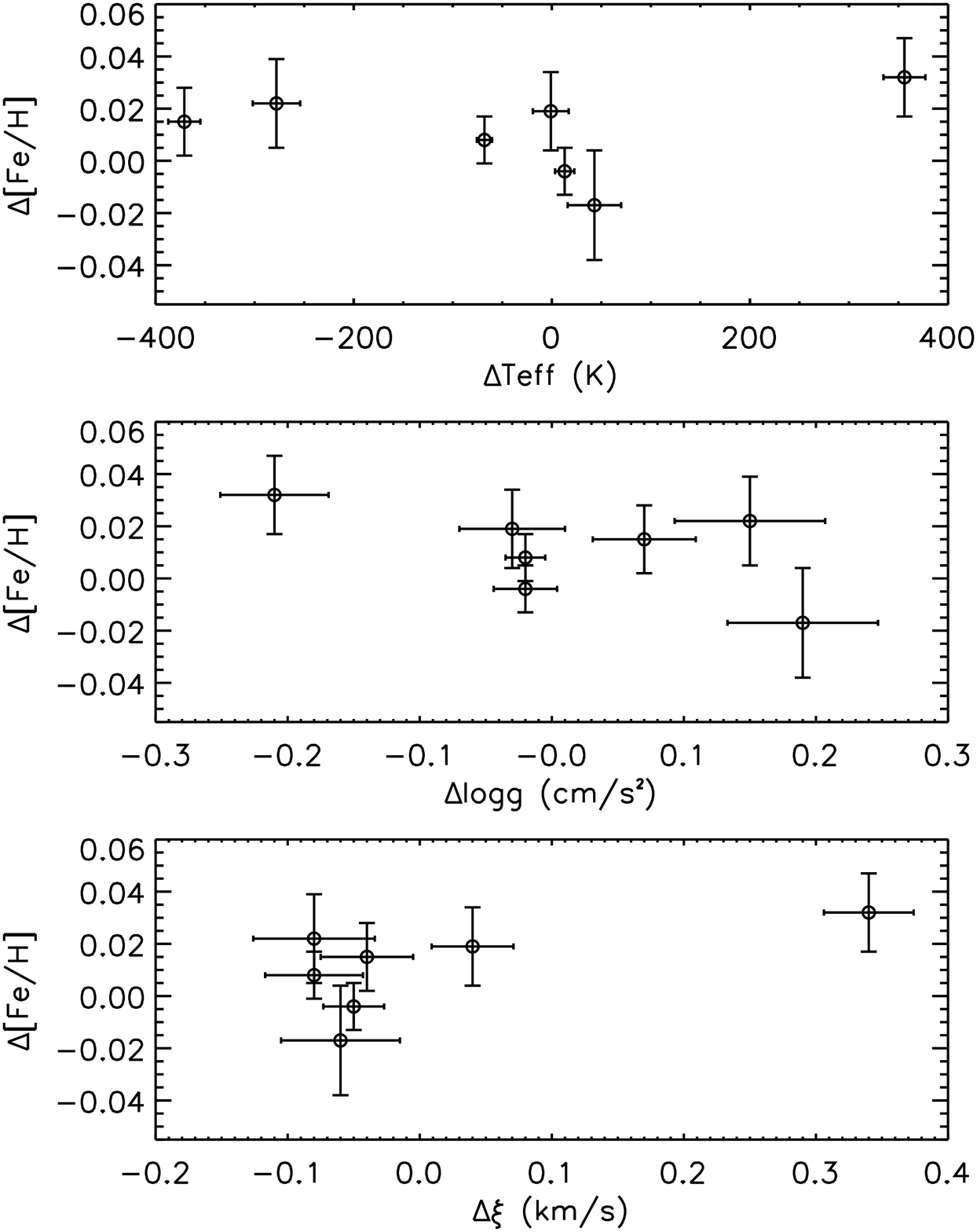}
\caption{Differential stellar parameters of 7 programme binary systems. Top panel: Differences in [Fe/H] as a function of differences in \teff. Middle panel: same as top panel but as a function of differences in \logg. Bottom panel: same as top panel but as a function of differences in \vt.}
\label{fig:dpara}
\end{figure}

We examined the correlation between differential stellar parameters in Figure \ref{fig:dpara}. There are no clear trends between $\Delta$[Fe/H] and other parameters ($\Delta$\teff, $\Delta$\logg\ and $\Delta$\vt). This demonstrates that the observed differences in [Fe/H] are not necessarily due to the differences in stellar temperature and gravity.

\begin{table*}
\caption{Adopted stellar atmospheric parameters of the programme binary systems.}
\centering
\label{t:para}
\begin{tabular}{@{}lcccc@{}}
\hline\hline
Object & $T_{\rm eff}$ & $\log g$ & [Fe/H] & $\xi_{\rm t}$ \\
 & (K) & (dex) & (dex) & (km/s) \\
\hline
HD\,106515A$^a$ & 5331 & 4.30 &  0.005 & 0.64 \\
HD\,106515B &  $-$68 $\pm$  8 & $-$0.02 $\pm$ 0.02 &  0.008 $\pm$ 0.009 & $-$0.08 $\pm$ 0.04 \\
HD\,108574$^a$  & 6175 & 4.28 &  0.092 & 1.23 \\
HD\,108575  & $-$371 $\pm$ 16 &  0.07 $\pm$ 0.04 &  0.015 $\pm$ 0.013 & $-$0.04 $\pm$ 0.04 \\
HD\,111484A$^a$ & 6134 & 4.20 &  0.048 & 1.20 \\
HD\,111484B &   $-$1 $\pm$ 18 & $-$0.03 $\pm$ 0.04 &  0.019 $\pm$ 0.015 &  0.04 $\pm$ 0.03 \\
HD\,133131A$^a$ & 5690 & 4.24 & $-$0.353 & 0.84 \\
HD\,133131B &   13 $\pm$ 10 & $-$0.02 $\pm$ 0.02 & $-$0.004 $\pm$ 0.009 & $-$0.05 $\pm$ 0.02 \\
HD\,146368A$^a$ & 6209 & 4.05 & $-$0.390 & 1.19 \\
HD\,146368B & $-$278 $\pm$ 24 &  0.15 $\pm$ 0.06 &  0.022 $\pm$ 0.017 & $-$0.08 $\pm$ 0.05 \\
HD\,216122A$^a$  & 6037 & 4.00 &  0.191 & 1.36 \\
HD\,216122B &   43 $\pm$ 27 &  0.19 $\pm$ 0.06 & $-$0.017 $\pm$ 0.021 & $-$0.06 $\pm$ 0.05 \\
HIP\,92961$^a$  & 6062 & 4.24 &  0.020 & 1.18 \\
HIP\,92962  &  356 $\pm$ 21 & $-$0.21 $\pm$ 0.04 &  0.032 $\pm$ 0.015 &  0.34 $\pm$ 0.03 \\
\hline
\end{tabular}
\\
$^a$ Adopted stellar parameters for the reference star.\\
\end{table*}

\subsection{Differential elemental abundances}

The chemical abundances of up to 29 elements in addition to Fe (C, O, Na, Mg, Al, Si, S, K, Ca, Sc, Ti, V, Cr, Mn, Co, Ni, Cu, Zn, Sr, Y, Zr, Ba, La, Ce, Nd, Sm, Eu, Gd and Dy) were derived for our objects, with the 1D LTE analysis and the curve of growth fitting from the spectral lines' EWs. The differential abundances of the 'B' component relative to the 'A' component of each binary system were derived with a strictly line-by-line approach, adopting the stellar parameters listed in Table \ref{t:para}. Hyperfine structure splitting (HFS) corrections were taken into account for V, Mn, Cu and Ba, with the HFS data taken from \citet{kb95} and \citet{bb15}. Following \citet{ama16}, we addressed 3D non-LTE corrections to the oxygen abundances for the 777\,nm triplet. The amount of differential 3D non-LTE abundance corrections for oxygen can vary between 0.01 - 0.08 dex for our programme binaries, which need to be taken into account. Except for oxygen, non-LTE effects for our differential abundances between two components of a binary system are almost negligible for available elements (see e.g., \citealt{liu18}), which should not alter our main results.

The differences in elemental abundances of our programme binaries are listed in Table \ref{t:abun}. The average abundance differences of all species for our programme binaries vary between 0.01 to 0.03 dex with a typical standard error $<$\,0.01 dex (standard deviation $<$\,0.02 dex). The uncertainties in differential abundances in Table \ref{t:abun} were calculated following the method employed by \citet{ben14}. They were derived by adding in quadrature the standard errors of the line-to-line scatter and the errors introduced from the stellar atmospheric parameters' uncertainties. The errors in the differential elemental abundances are below 0.02 dex for most species (see Table \ref{t:abun}). We note the average error in differential abundances ($\Delta$[X/H]) for all of our programme binaries are only $\lesssim$ 0.015 dex (3.5\%). The precision of the abundance analysis was improved by 2 - 5 times when compared to the previous studies, as discussed in the section below.

\begin{table*}
\caption{Differential elemental abundances of the programme binary systems.}
\centering
\label{t:abun}
\begin{tabular}{@{}lrrrrrrrrrrrrrr@{}}
\hline\hline
Species & $\Delta$[X/H] & $\Delta$[X/H] & $\Delta$[X/H] & $\Delta$[X/H] & $\Delta$[X/H] & $\Delta$[X/H] & $\Delta$[X/H] \\
  & HD106515B$-$A$^a$ & HD108575$-$74 & HD111484B$-$A & HD133131B$-$A$^a$ & HD146368B$-$A & HD216122B$-$A & HIP92962$-$61 \\
\hline
C\,{\sc i} & 0.013 $\pm$ 0.032 & 0.060 $\pm$ 0.021 & $-$0.023 $\pm$ 0.026 & ... & 0.054 $\pm$ 0.009 & $-$0.004 $\pm$ 0.021 & $-$0.047 $\pm$ 0.011 \\
O\,{\sc i}$^b$ & $-$0.010 $\pm$ 0.015 & 0.072 $\pm$ 0.017 & 0.004 $\pm$ 0.011 & $-$0.030 $\pm$ 0.015 & 0.088 $\pm$ 0.017 & $-$0.024 $\pm$ 0.026 & $-$0.019 $\pm$ 0.012 \\
Na\,{\sc i} & 0.018 $\pm$ 0.011 & 0.006 $\pm$ 0.012 & 0.013 $\pm$ 0.009 & $-$0.007 $\pm$ 0.008 & 0.009 $\pm$ 0.008 & $-$0.031 $\pm$ 0.010 & 0.056 $\pm$ 0.014 \\
Mg\,{\sc i} & 0.000 $\pm$ 0.005 & 0.035 $\pm$ 0.009 & 0.034 $\pm$ 0.005 & $-$0.015 $\pm$ 0.005 & 0.044 $\pm$ 0.010 & $-$0.005 $\pm$ 0.025 & 0.032 $\pm$ 0.013 \\
Al\,{\sc i} & 0.004 $\pm$ 0.009 & 0.051 $\pm$ 0.016 & 0.047 $\pm$ 0.003 & $-$0.014 $\pm$ 0.011 & 0.066 $\pm$ 0.019 & $-$0.001 $\pm$ 0.014 & 0.028 $\pm$ 0.012 \\
Si\,{\sc i} & 0.008 $\pm$ 0.007 & 0.004 $\pm$ 0.006 & 0.033 $\pm$ 0.007 & $-$0.012 $\pm$ 0.007 & 0.004 $\pm$ 0.009 & $-$0.030 $\pm$ 0.010 & 0.045 $\pm$ 0.008 \\
S\,{\sc i} & 0.019 $\pm$ 0.029 & 0.018 $\pm$ 0.029 & 0.034 $\pm$ 0.016 & $-$0.017 $\pm$ 0.025 & 0.053 $\pm$ 0.026 & $-$0.022 $\pm$ 0.023 & 0.029 $\pm$ 0.017 \\
K\,{\sc i} & $-$0.015 $\pm$ 0.026 & ... & $-$0.002 $\pm$ 0.028 & 0.015 $\pm$ 0.026 & ... & $-$0.032 $\pm$ 0.029 & ... \\
Ca\,{\sc i} & 0.021 $\pm$ 0.009 & 0.048 $\pm$ 0.011 & 0.017 $\pm$ 0.011 & $-$0.011 $\pm$ 0.007 & 0.018 $\pm$ 0.013 & $-$0.001 $\pm$ 0.017 & 0.050 $\pm$ 0.013 \\
Sc\,{\sc ii} & 0.012 $\pm$ 0.015 & 0.054 $\pm$ 0.020 & 0.023 $\pm$ 0.017 & $-$0.026 $\pm$ 0.013 & 0.028 $\pm$ 0.026 & 0.000 $\pm$ 0.028 & 0.039 $\pm$ 0.021 \\
Ti\,{\sc i} & 0.022 $\pm$ 0.009 & 0.031 $\pm$ 0.016 & 0.031 $\pm$ 0.016 & $-$0.020 $\pm$ 0.011 & $-$0.014 $\pm$ 0.022 & $-$0.010 $\pm$ 0.024 & 0.045 $\pm$ 0.015 \\
Ti\,{\sc ii} & 0.003 $\pm$ 0.009 & 0.048 $\pm$ 0.016 & 0.049 $\pm$ 0.017 & $-$0.012 $\pm$ 0.011 & 0.028 $\pm$ 0.022 & $-$0.011 $\pm$ 0.029 & 0.035 $\pm$ 0.019 \\
V\,{\sc i} & 0.054 $\pm$ 0.017 & 0.053 $\pm$ 0.021 & 0.024 $\pm$ 0.020 & $-$0.010 $\pm$ 0.019 & 0.012 $\pm$ 0.026 & 0.004 $\pm$ 0.026 & 0.060 $\pm$ 0.020 \\
Cr\,{\sc i} & 0.021 $\pm$ 0.008 & 0.039 $\pm$ 0.014 & 0.015 $\pm$ 0.014 & $-$0.012 $\pm$ 0.008 & 0.028 $\pm$ 0.016 & $-$0.042 $\pm$ 0.020 & 0.033 $\pm$ 0.014 \\
Cr\,{\sc ii} & $-$0.002 $\pm$ 0.013 & ... & $-$0.007 $\pm$ 0.015 & $-$0.016 $\pm$ 0.009 & 0.028 $\pm$ 0.019 & ... & ... \\
Mn\,{\sc i} & 0.012 $\pm$ 0.010 & 0.028 $\pm$ 0.018 & 0.018 $\pm$ 0.017 & $-$0.013 $\pm$ 0.011 & ... & 0.009 $\pm$ 0.023 & $-$0.012 $\pm$ 0.026 \\
Fe\,{\sc i} & 0.007 $\pm$ 0.006 & 0.015 $\pm$ 0.010 & 0.019 $\pm$ 0.011 & $-$0.004 $\pm$ 0.006 & 0.023 $\pm$ 0.013 & $-$0.017 $\pm$ 0.016 & 0.033 $\pm$ 0.012 \\
Fe\,{\sc ii} & 0.007 $\pm$ 0.011 & 0.015 $\pm$ 0.014 & 0.019 $\pm$ 0.017 & $-$0.004 $\pm$ 0.010 & 0.022 $\pm$ 0.019 & $-$0.018 $\pm$ 0.024 & 0.033 $\pm$ 0.018 \\
Co\,{\sc i} & 0.035 $\pm$ 0.016 & 0.030 $\pm$ 0.025 & $-$0.012 $\pm$ 0.019 & $-$0.037 $\pm$ 0.010 & 0.043 $\pm$ 0.016 & $-$0.053 $\pm$ 0.023 & 0.024 $\pm$ 0.024 \\
Ni\,{\sc i} & 0.008 $\pm$ 0.006 & 0.021 $\pm$ 0.011 & 0.046 $\pm$ 0.013 & $-$0.012 $\pm$ 0.007 & 0.043 $\pm$ 0.015 & $-$0.009 $\pm$ 0.018 & 0.018 $\pm$ 0.013 \\
Cu\,{\sc i} & 0.044 $\pm$ 0.008 & 0.042 $\pm$ 0.027 & 0.001 $\pm$ 0.011 & $-$0.002 $\pm$ 0.020 & 0.046 $\pm$ 0.027 & $-$0.012 $\pm$ 0.015 & ... \\
Zn\,{\sc i} & 0.028 $\pm$ 0.011 & 0.001 $\pm$ 0.024 & $-$0.009 $\pm$ 0.013 & $-$0.012 $\pm$ 0.032 & 0.010 $\pm$ 0.013 & $-$0.018 $\pm$ 0.021 & $-$0.024 $\pm$ 0.016 \\
Sr\,{\sc i} & ... & 0.023 $\pm$ 0.030 & 0.030 $\pm$ 0.029 & 0.005 $\pm$ 0.026 & $-$0.011 $\pm$ 0.031 & 0.041 $\pm$ 0.033 & 0.061 $\pm$ 0.029 \\
Y\,{\sc ii} & 0.039 $\pm$ 0.020 & 0.048 $\pm$ 0.041 & 0.029 $\pm$ 0.022 & $-$0.020 $\pm$ 0.012 & 0.020 $\pm$ 0.024 & 0.004 $\pm$ 0.030 & 0.022 $\pm$ 0.031 \\
Zr\,{\sc ii} & 0.052 $\pm$ 0.030 & ... & 0.036 $\pm$ 0.032 & 0.000 $\pm$ 0.028 & $-$0.021 $\pm$ 0.035 & ... & ... \\
Ba\,{\sc ii} & 0.009 $\pm$ 0.013 & 0.040 $\pm$ 0.019 & $-$0.003 $\pm$ 0.018 & 0.002 $\pm$ 0.010 & ... & $-$0.014 $\pm$ 0.028 & 0.032 $\pm$ 0.024 \\
La\,{\sc ii} & $-$0.011 $\pm$ 0.026 & 0.044 $\pm$ 0.032 & 0.029 $\pm$ 0.032 & 0.004 $\pm$ 0.028 & 0.010 $\pm$ 0.033 & 0.032 $\pm$ 0.038 & 0.047 $\pm$ 0.044 \\
Ce\,{\sc ii} & 0.010 $\pm$ 0.021 & 0.058 $\pm$ 0.024 & $-$0.006 $\pm$ 0.022 & $-$0.018 $\pm$ 0.017 & 0.029 $\pm$ 0.037 & 0.001 $\pm$ 0.050 & 0.011 $\pm$ 0.024 \\
Nd\,{\sc ii} & $-$0.002 $\pm$ 0.026 & $-$0.015 $\pm$ 0.036 & 0.003 $\pm$ 0.032 & $-$0.036 $\pm$ 0.028 & 0.015 $\pm$ 0.039 & ... & 0.034 $\pm$ 0.037 \\
Sm\,{\sc ii} & 0.059 $\pm$ 0.013 & 0.003 $\pm$ 0.035 & 0.008 $\pm$ 0.035 & $-$0.030 $\pm$ 0.012 & $-$0.009 $\pm$ 0.037 & 0.034 $\pm$ 0.037 & 0.051 $\pm$ 0.031 \\
Eu\,{\sc ii} & 0.024 $\pm$ 0.026 & ... & 0.024 $\pm$ 0.030 & ... & ... & ... & 0.015 $\pm$ 0.031 \\
Gd\,{\sc ii} & 0.034 $\pm$ 0.027 & 0.039 $\pm$ 0.032 & 0.036 $\pm$ 0.032 & $-$0.024 $\pm$ 0.028 & ... & ... & ... \\
Dy\,{\sc ii} & $-$0.006 $\pm$ 0.026 & ... & ... & 0.005 $\pm$ 0.028 & ... &  ... &  ... \\
\hline
\end{tabular}
{\raggedright $^a$ Planet-hosting pairs.\\ \par}
{\raggedright $^b$ Differential 3D non-LTE abundance corrections applied.\\ \par}
\end{table*}

\subsection{Comparison to the previous studies}

We compared the stellar atmospheric parameters of two common binaries (HD\,106515A/B and HD\,133131A/B) from this work to that from \citet{des04} and \citet{des06}. The differences are: 1 $\pm$ 60 K for $\Delta$\teff, 0.00 $\pm$ 0.07 dex for $\Delta$\logg, and $-$0.013 $\pm$ 0.039 dex for $\Delta$[Fe/H]. In addition, we compared the stellar parameters of HD\,106515A/B between this work to that of S19. Between the two components, our results show less differences in \teff\ and \logg, but very similar [Fe/H] as in S19. We also compared our results to T16's for another pair HD\,133131A/B. The stellar parameters agree well with these studies within the uncertainties. We note that the typical uncertainties in this work are about 5 times better than that in \citet{des04,des06}, 2 - 3 times better than that in S19, and 50\% better then that in T16. The better quality spectra that lead to more accurate measurements might be the major reason of the more precise results in this study. Meanwhile the critical selection of lines (avoiding potentially blended lines and strong lines) might as well help yield stellar parameters more precisely.

\begin{figure}
\centering
\includegraphics[width=\columnwidth]{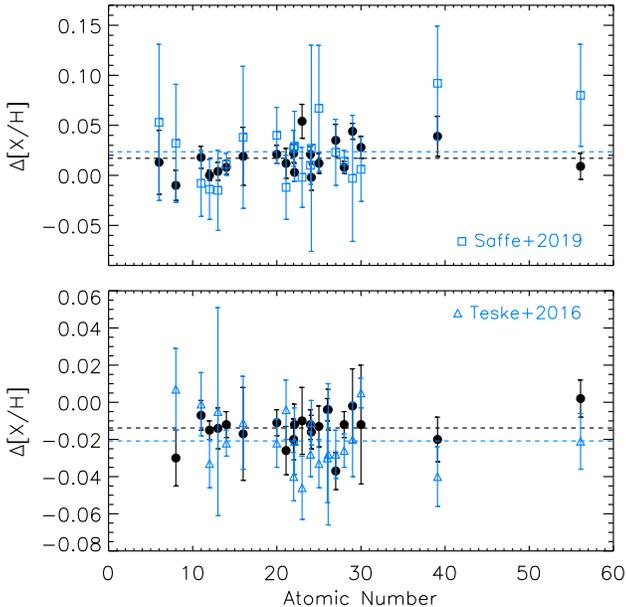}
\caption{The differential abundances from this work in comparison to the previous studies. Top panel: Differences in abundances $\Delta$[X/H]\,(HD\,106515B$-$HD\,106515A) for common elements in this work (black circles) and S19's work (blue rectangles). The dashed lines mark out the mean values of abundance differences. Bottom panel: Similar as in the top panel, but comparing $\Delta$[X/H]\,(HD\,133131B$-$HD\,133131A) between this work (black circles) and T16's work (blue triangles).}
\label{fig:comp}
\end{figure}

In Figure \ref{fig:comp} we compare the differential abundances for HD\,106515A/B between this work and S19, and for HD\,133131A/B between this work and T16\footnote{\citet{des04} and \citet{des06} only derived stellar parameters, not detailed abundances and hence no such comparison is possible.} For the elements in common, the average abundance differences are 0.007 $\pm$ 0.034 dex for HD\,106515B (with relative to HD\,106515A) when comparing this work to S19. Similarly, the average abundance differences are $-$0.007 $\pm$ 0.018 dex for HD\,133131B (with relative to HD\,133131A) when comparing our results to T16's. The average uncertainties in the differential abundances are only 0.012 dex and 0.013 dex for HD\,133131A/B and HD\,106515A/B, which are about 1.5 - 3 times smaller than that in T16 and S19 (0.018 dex and 0.043 dex, respectively). This emphasizes the significance of higher quality spectra.

\section{Discussion}

\subsection{Relative elemental abundances and condensation temperature trends}

\begin{figure}
\centering
\includegraphics[width=\columnwidth]{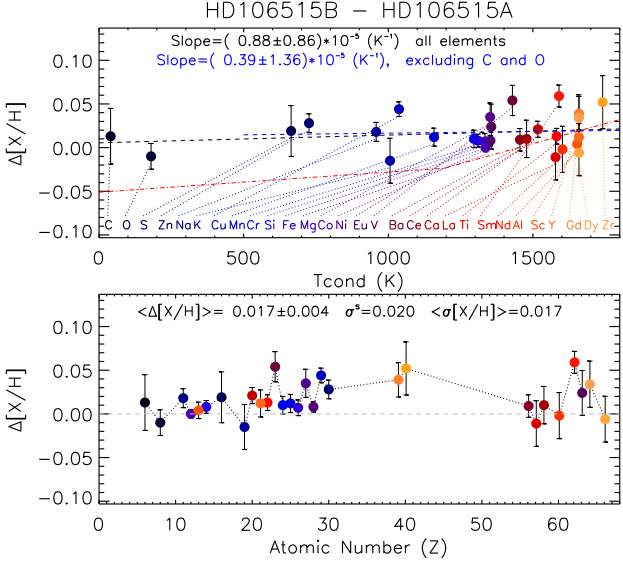}
\caption{Differential elemental abundances of HD\,106515B relative to HD\,106515A, in which HD\,106515A hosts a giant planet. The elements are colour coded with \tcond. Top panel: Differential elemental abundances of HD\,106515B relative to HD\,106515A versus \tcond. The dashed lines represent the linear least-squares fits to the data (black: including C and O; blue: excluding C and O). The red dash-dot line represents the \tcond\ trend from M09. Bottom panel: Differential elemental abundances of HD\,106515B relative to HD\,106515A versus atomic number (Z) for all the elements.}
\label{fig:tcond1}
\end{figure}

\begin{figure}
\centering
\includegraphics[width=\columnwidth]{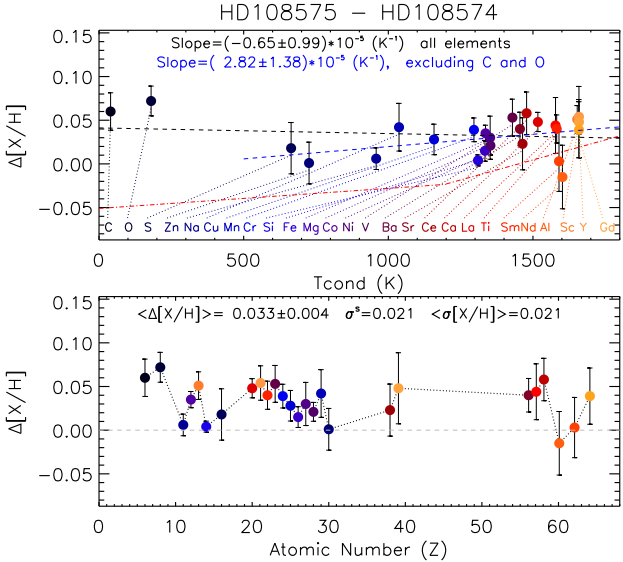}
\caption{Same as Figure \ref{fig:tcond1} but for HD\,108575 relative to HD\,108574.}
\label{fig:tcond2}
\end{figure}

\begin{figure}
\centering
\includegraphics[width=\columnwidth]{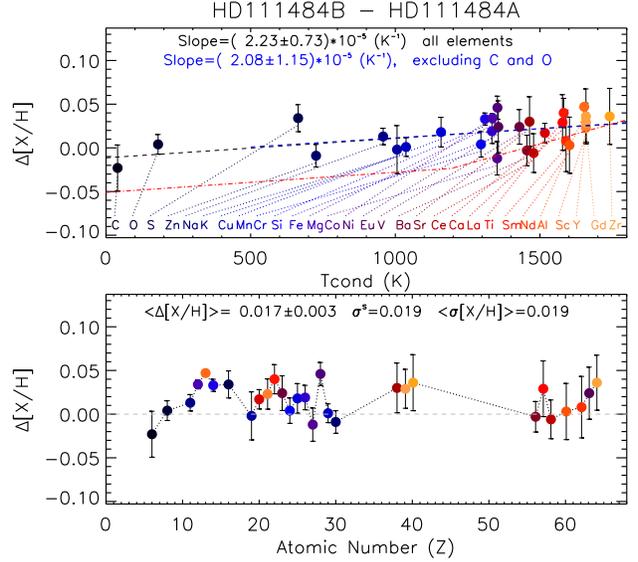}
\caption{Same as Figure \ref{fig:tcond1} but for HD\,111484B relative to HD\,111484A.}
\label{fig:tcond3}
\end{figure}

\begin{figure}
\centering
\includegraphics[width=\columnwidth]{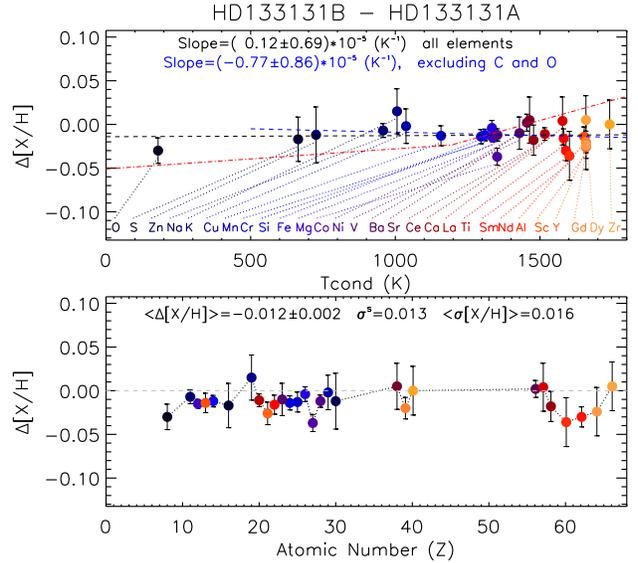}
\caption{Same as Figure \ref{fig:tcond1} but for HD\,133131B relative to HD\,133131A, in which both HD\,133131A and HD\,133131B host giant planets.}
\label{fig:tcond4}
\end{figure}

\begin{figure}
\centering
\includegraphics[width=\columnwidth]{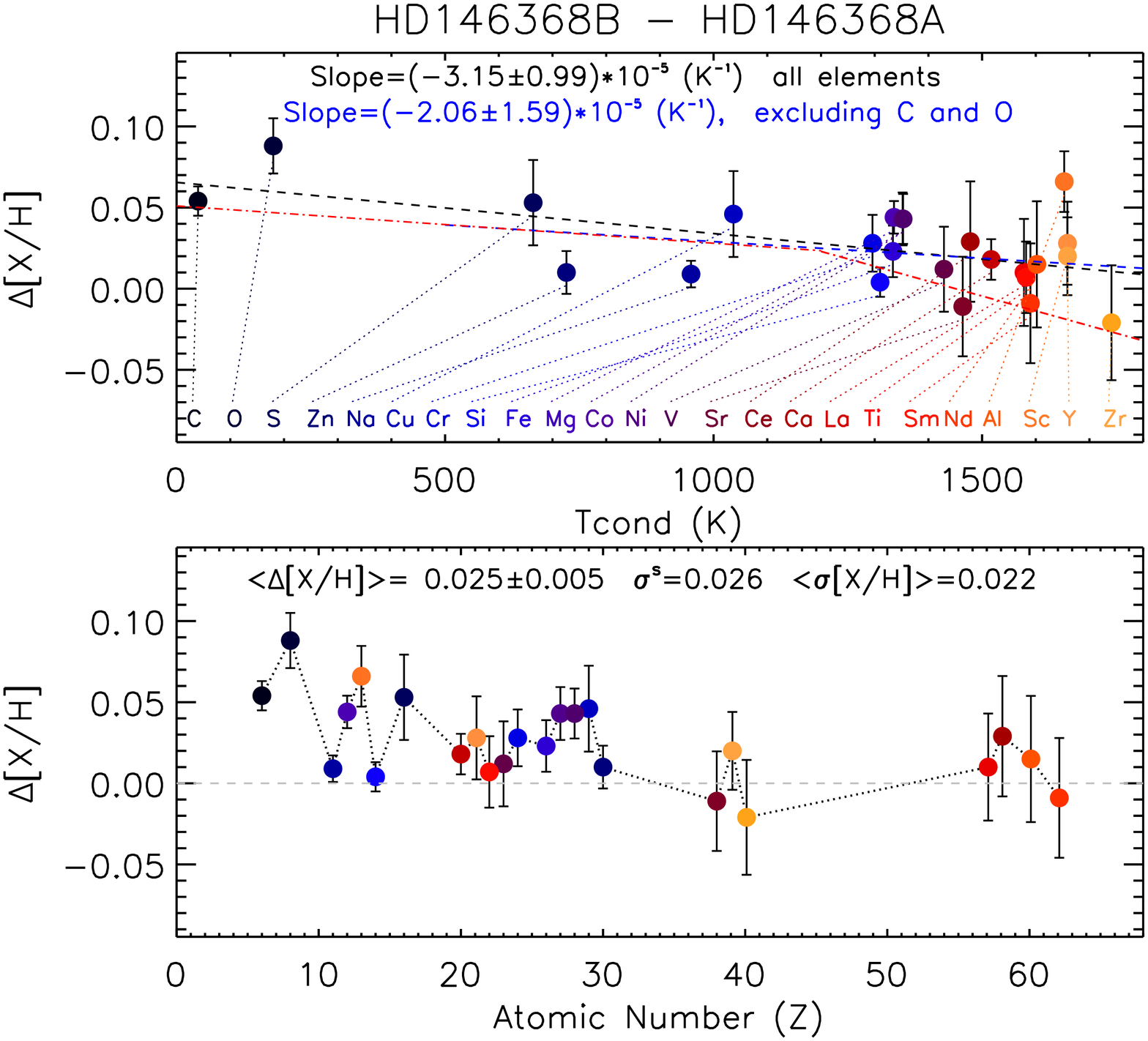}
\caption{Same as Figure \ref{fig:tcond1} but for HD\,146368B relative to HD\,146368A. The \tcond\ trend from M09 has been switched to the opposite sign.}
\label{fig:tcond5}
\end{figure}

\begin{figure}
\centering
\includegraphics[width=\columnwidth]{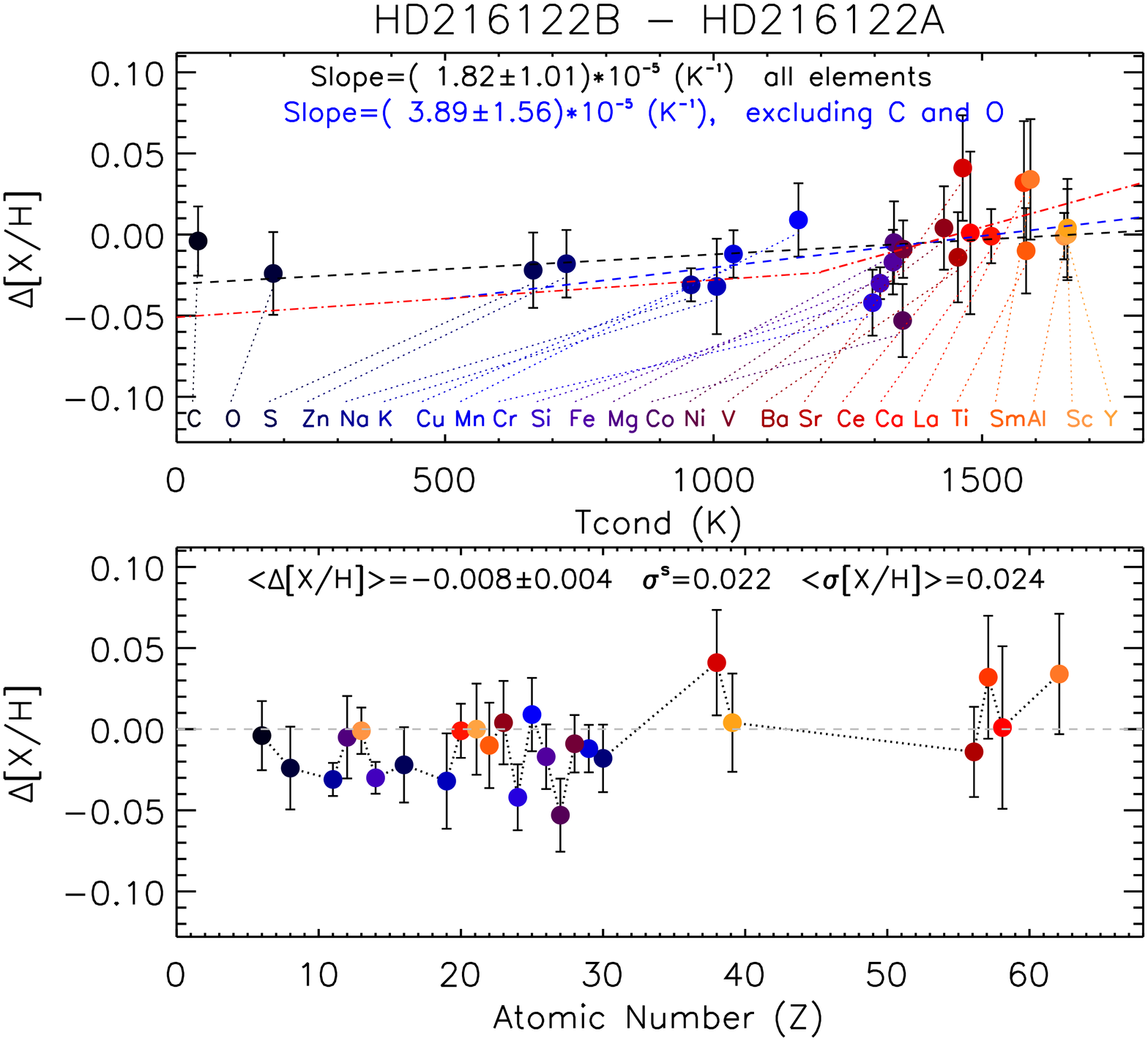}
\caption{Same as Figure \ref{fig:tcond1} but for HD\,216122B relative to HD\,216122A.}
\label{fig:tcond6}
\end{figure}

\begin{figure}
\centering
\includegraphics[width=\columnwidth]{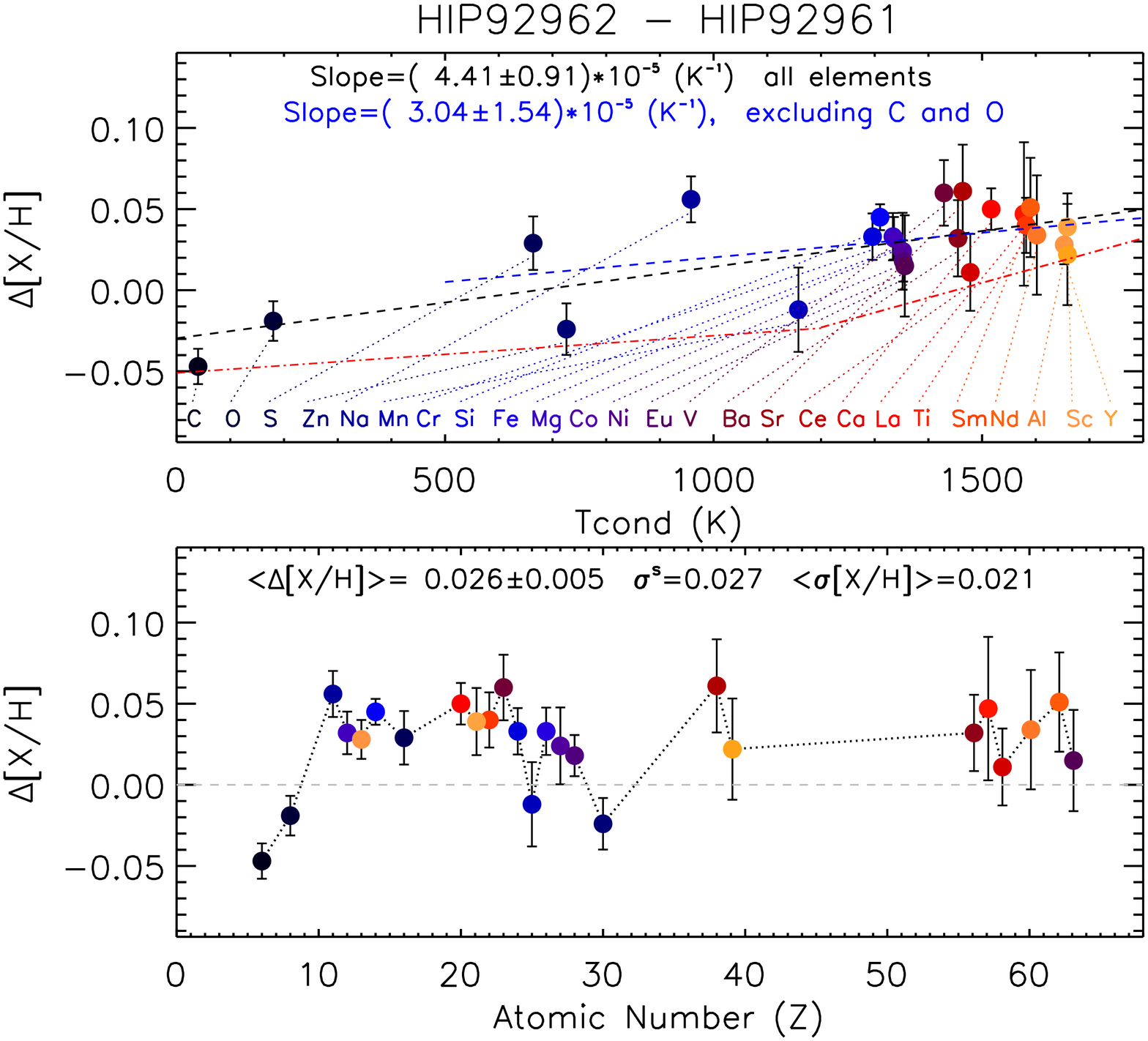}
\caption{Same as Figure \ref{fig:tcond1} but for HIP\,92962 relative to HIP\,92961.}
\label{fig:tcond7}
\end{figure}

The differential elemental abundances of the programme binary systems as a function of atomic number and condensation temperature were shown in Figure \ref{fig:tcond1} - \ref{fig:tcond7}. We note the differences in neutron-capture element abundances for a given binary pair exhibits no clear correlation with the $r$-process element fractions in the solar system \citep{bis11}. The lack of correlation would suggest that neutron-capture element abundance differences cannot be attributed solely to either the $s$- nor $r$-process, which is different for the case of 18\,Sco \citep{mel14}.

The correlation between elemental abundances and \tcond\footnote{The \tcond\ values (50\% condensation) were taken from \citet{lod03}.}, namely \tcond\ trends, can be attributed to planet formation processes as introduced before. Table \ref{t:stat} lists the overall abundance differences ($<$$\Delta$[X/H]$>$), the corresponding standard error, the standard deviation ($\sigma^s$) and the average error ($<$$\sigma$[X/H]$>$), as well as the slopes of \tcond\ trends for our programme binaries. We found that among the 7 pairs, 3 of them (HD\,106515A/B, HD\,108574/75 and HD\,133131A/B) have almost zero slopes of \tcond\ trends ($< 1.0 \times 10^{-5}$ K$^{-1}$) with overall abundance differences of 0.01 - 0.03 dex, including 2 pairs hosting known giant planets (HD\,106515A/B and HD\,133131A/B). HD\,216122A/B shows non-significant slope of \tcond\ trend of $(1.82 \pm 1.01) \times 10^{-5}$ K$^{-1}$. Meanwhile the other 3 pairs (HD\,111484A/B, HD\,146368A/B and HIP92961/62) exhibit slopes of \tcond\ trends of $(2.23 \pm 0.73) \times 10^{-5}$ K$^{-1}$, $(-3.15 \pm 0.99) \times 10^{-5}$ K$^{-1}$ and $(4.41 \pm 0.91) \times 10^{-5}$ K$^{-1}$, with the significance level of 3.1\,$\sigma$, 3.2\,$\sigma$ and 4.8\,$\sigma$, respectively. 

In addition, we are aware that the \tcond\ trends can be driven by C and O for some cases. Therefore we also applied the linear square fits to the data excluding C and O. We found that HD\,108574/85 shows a \tcond-slope with 2.0\,$\sigma$ significance instead of zero slope, and the \tcond-slope of HD\,216122A/B becomes more significant (2.5\,$\sigma$). Meanwhile the slopes of \tcond\ trends of HD\,111484A/B, HD\,146368A/B and HIP\,92961/62 become less significant (1.8\,$\sigma$, 1.3\,$\sigma$ and 2.0\,$\sigma$). The results indicate that C and O play a key role in determining the \tcond\ trends. In our following analysis, we still adopt the fitting results including C and O.

The volatile-to-refractory abundance ratios ([Vol/Ref]) of the programme binaries are derived by taking the mean for each group of elements\footnote{The \tcond\ value that separates volatiles from refractories is defined as 1100\,K.} and listed in Table \ref{t:stat}. Although the binary systems showing visible \tcond\ trends naturally differ most in their [Vol/Ref], the results are not significant enough for us to draw a conclusion, given some might be driven by C and O as stated above. Below we mainly discuss the potential hypotheses of planet formation that can induce the observed differences in elemental abundances and the corresponding \tcond\ trends within a given pair of binary stars.

\subsection{Effects of planet formation}

M09 has reported that, in comparison to the majority of solar twins, the Sun is depleted by 20\% in refractory (high \tcond) elements. They proposed the scenario that the formation of rocky planets in early stage takes away from the proto-planetary disc more refractory material than volatile material, leading to the observed differences between refractories and volatiles in a planet-hosting star. A recent study has attributed the observed abundance pattern of the Sun to the formation of its distant giant planets \citep{bo20}, as they can trap significant amounts of dust exterior to their orbits, shutting out the accretion of the dust into the Sun, thereby leading to the lower refractory abundances in the Sun.

Following the scenario that planet formation takes away material from the star, one would expect the star hosting more massive planets to be more metal-poor in refractory elements. Two pairs in our sample (HD\,106515A/B and HD\,133131A/B) host known planets. HD\,106515A hosts a giant planet with minimum mass of 9.6\,\mjup\ and semi-major axis of 4.6\,AU. This pair has an overall abundance difference of 0.017 $\pm$ 0.004 dex ($\sigma^s$ = 0.020 dex) with almost zero slope $(0.88 \pm 0.86) \times 10^{-5}$ K$^{-1}$ of \tcond\ trend. The 'A' component is marginally more metal-poor than the 'B' component without clear \tcond\ trend, despite hosting a distant, very massive planet. For HD\,133131A/B, the 'A' component hosts two planets b and c with minimum masses of 1.43\,\mjup\ and 0.48\,\mjup; the 'B' component hosts a planet with minimum mass of 2.5\,\mjup. All of these planets are further away than 1\,AU. This pair has an overall abundance difference of $-$0.012 $\pm$ 0.002 dex ($\sigma^s$ = 0.013 dex) with also zero slope $(0.12 \pm 0.69) \times 10^{-5}$ K$^{-1}$ of \tcond\ trend. The 'B' component, which hosts more massive planet(s), is marginally more metal-poor than the 'A' component. We note that the results, although fulfill the expectation of the sign of abundance differences, can not be fully explained by the hypotheses proposed by either M09 or \citet{bo20}, due to the lack of visible \tcond\ trends and the fact that the other 1 - 2 pairs without known planet also exhibit similar abundance pattern. Indeed, our results might imply that formation of (gas) giant planet does not necessarily imprint chemical signatures and the mass of planet does not necessarily produce the corresponding scale of abundance differences. Although we remind the readers it is highly probable that some pairs in our sample do host unknown planet(s).

Furthermore, it has recently been proposed that planet formation location might influence the abundance pattern of planet-hosting stars (i.e. \tcond\ trends). The theoretical simulations in \citet{bit18} demonstrated that a planet formed inside e.g., H$_2$O/CO ice-line can produce much larger differences in its host stars' surface abundance ratios (e.g., [Fe/O] or [Fe/C]) than for planets formed out of the H$_2$O/CO ice-line. At the disk location beyond the major volatile snow lines, the disk heavy elements are in the solid phase. The accreted material onto planets has the same refractory-to-volatile ratio compared to the solar value, and therefore, the planet growth has no contribution on the \tcond\ trend. On the other hand, when the planets form interior to the H$_2$O/CO snow lines, the accreted core is volatile-depleted, and the host star would reveal a high abundance of volatile elements compared to the refractory ones. As a consequence, the planet formation location plays a decisive role in determining the observed abundance pattern of binary systems.

Following this hypothesis, whether we are able to detect significant \tcond\ trends could depend on the formation location of potential planets. From our results, 3 of 7 pairs (43\%) have marginal abundance differences but without \tcond\ trends; while 3 or 4 of 7 pairs (43\% or 57\%) show abundance differences with visible \tcond\ trends. The variety in the observed abundance pattern in our sample might arise from the variations of planet formation location (e.g., inside or outside the H$_2$O/CO ice-line). However, the small number of sample and the lack of planet information complicate the situation and limit our capability to draw a solid conclusion.

\subsection{Effects of atomic diffusion}

Atomic diffusion is the combination effect of two processes -- gravitational settling and radiative acceleration \citet{mic84,mic15,dot17} -- that can alter the surface elemental abundances of a star depending on its evolutionary stage. Binaries are assumed to form together and share the same initial chemical composition. However, for a pair of binary stars with different stellar atmospheric parameters (\teff\ and \logg), their surface elemental abundance may still be altered differently when they evolve. The signatures of atomic diffusion were previously detected in the globular clusters \citet{kor06,nor12,gru16} and the open cluster M67 \citet{gao18,liu19,sou19}, but have yet to be established in binary systems. 

In our sample, the effects of atomic diffusion can be revealed because a few pairs have relative large differences in their \teff\ and \logg, and the abundance precision in this study is sufficiently high. Indeed, we suspect the observed abundance offsets in some binaries may be due to atomic diffusion. Therefore we examine the atomic diffusion effects, especially for those pairs with relative large \logg\ differences.

\begin{figure*}
\centering
\includegraphics[width=\textwidth]{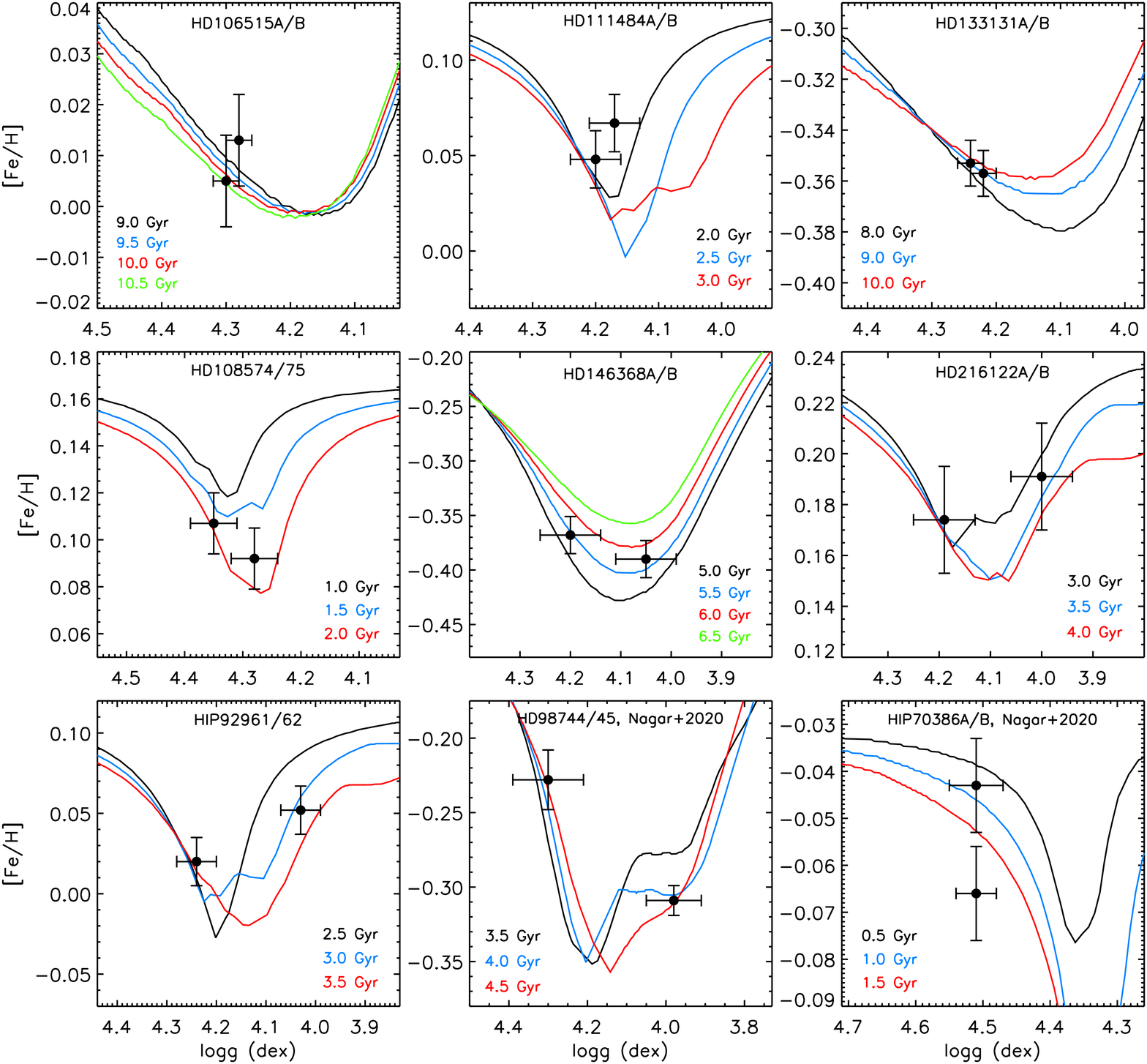}
\caption{Predicted change in [Fe/H] against \logg\ with atomic diffusion taken into account \citep{cho16,dot17} for the seven binaries in our sample, as well as two additional binaries from \citet{nag20}. The first three pairs in the top panel have \logg\ difference $<$\,0.05 dex, A, in which HD\,106515A/B and HD\,133131A/B host planets.} The next four pairs have relative large difference in \logg\ (HD\,108574/75, HD\,146368A/B, HD216122A/B and HIP92961/62). The coloured solid lines represent the model predictions for different ages as labelled. The observational results are shown as filled circles.
\label{fig:atm}
\end{figure*}

We compare the observed results with the stellar evolutionary models for each pair with its average metallicity and corresponding age range. The adopted models are taken from \citet{cho16} and \citet{dot17}\footnote{\url{http://waps.cfa.harvard.edu/MIST/}.}, where overshooting mixing, turbulent and atomic diffusion were taken into account. Figure \ref{fig:atm} shows the expected changes in [Fe/H] against \logg\ of all the pairs, and how they compare with our observed metallicity offsets. We also include two pairs (HD\,98744/45 and HIP\,70386A/B) from \citet{nag20} for examination, because both pairs show overall abundance offsets but no \tcond\ trends. We found that the observed differences in [Fe/H] agree with the theoretical models for four of our pairs (HD\,108574/75, HD\,146368A/B, HD216122A/B and HIP\,92961/62) and one pair (HD\,98744/45) from \citet{nag20}, all with \logg\ differences exceeding 0.05 dex.

\begin{figure*}
\centering
\includegraphics[width=\textwidth]{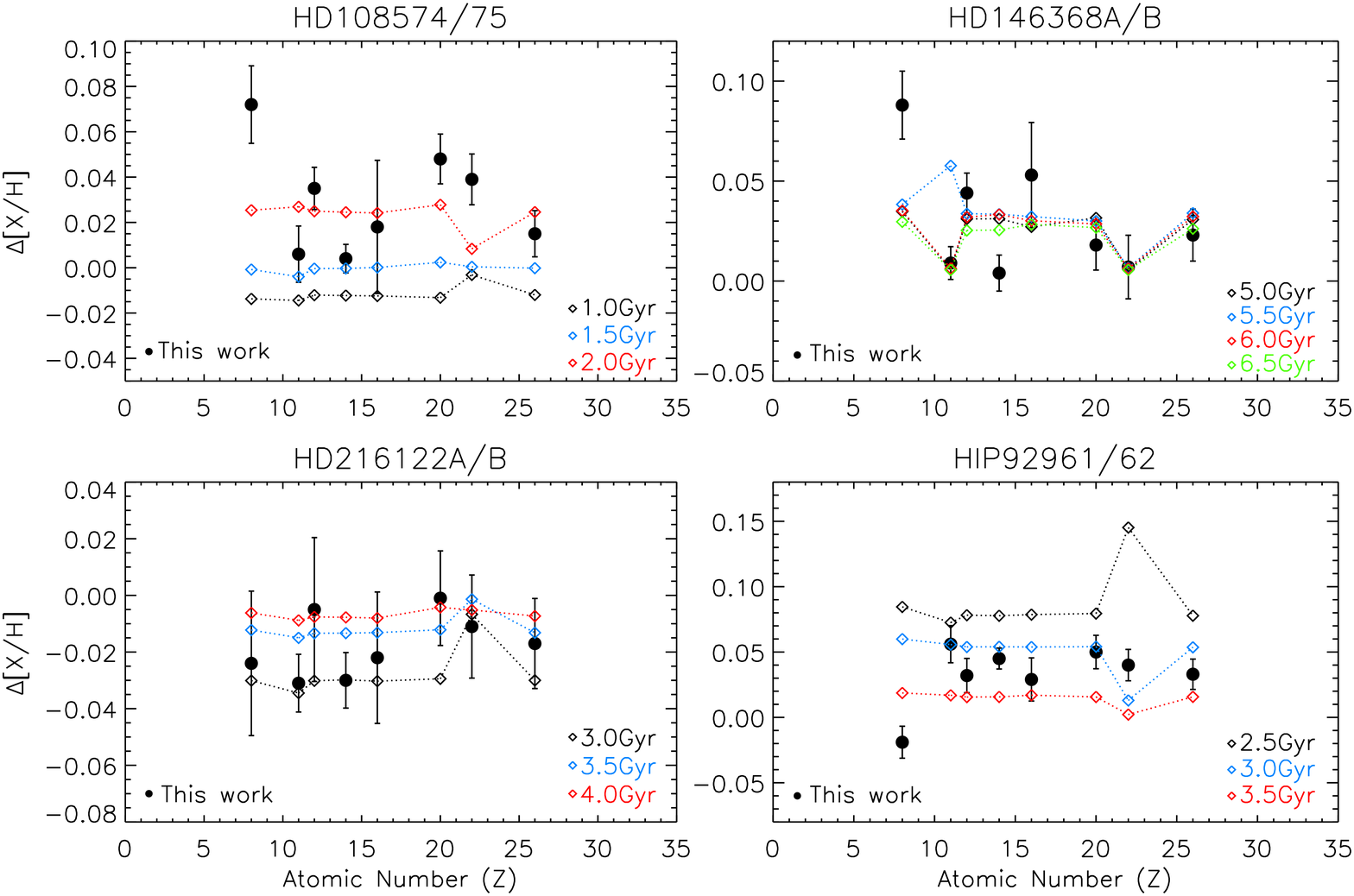}
\caption{Predicted change in [X/H] against atomic numbers (Z) with atomic diffusion taken into account \citep{cho16,dot17} for four pairs of binary stars with relative large difference in \logg\ (HD\,108574/75, HD\,146368A/B, HD\,216122A/B and HIP\,92961/62). The coloured symbols represent the model predictions for different ages as labelled. The observational results are shown as filled circles.}
\label{fig:bbb}
\end{figure*}

We then compare the observed abundance differences within each of our four pairs against the model predictions for 8 elements (O, Na, Mg, Si, S, Ca, Ti and Fe) in Figure \ref{fig:bbb}. We found that the observed differences in [X/H] (at a level of 0.02\,dex) agree with the theoretical models qualitatively at the estimated age range. This indicates that the overall abundance offsets for these four pairs could be due to the signatures of atomic diffusion, rather than planet formation. Our results emphasize that the effects of atomic diffusion are non-negligible, even in binaries, when the two components are not similar enough. They should be taken into account and examined in high-precision spectroscopic studies. 

\subsection{Abundance differences and binary separation}

\begin{figure}
\centering
\includegraphics[width=\columnwidth]{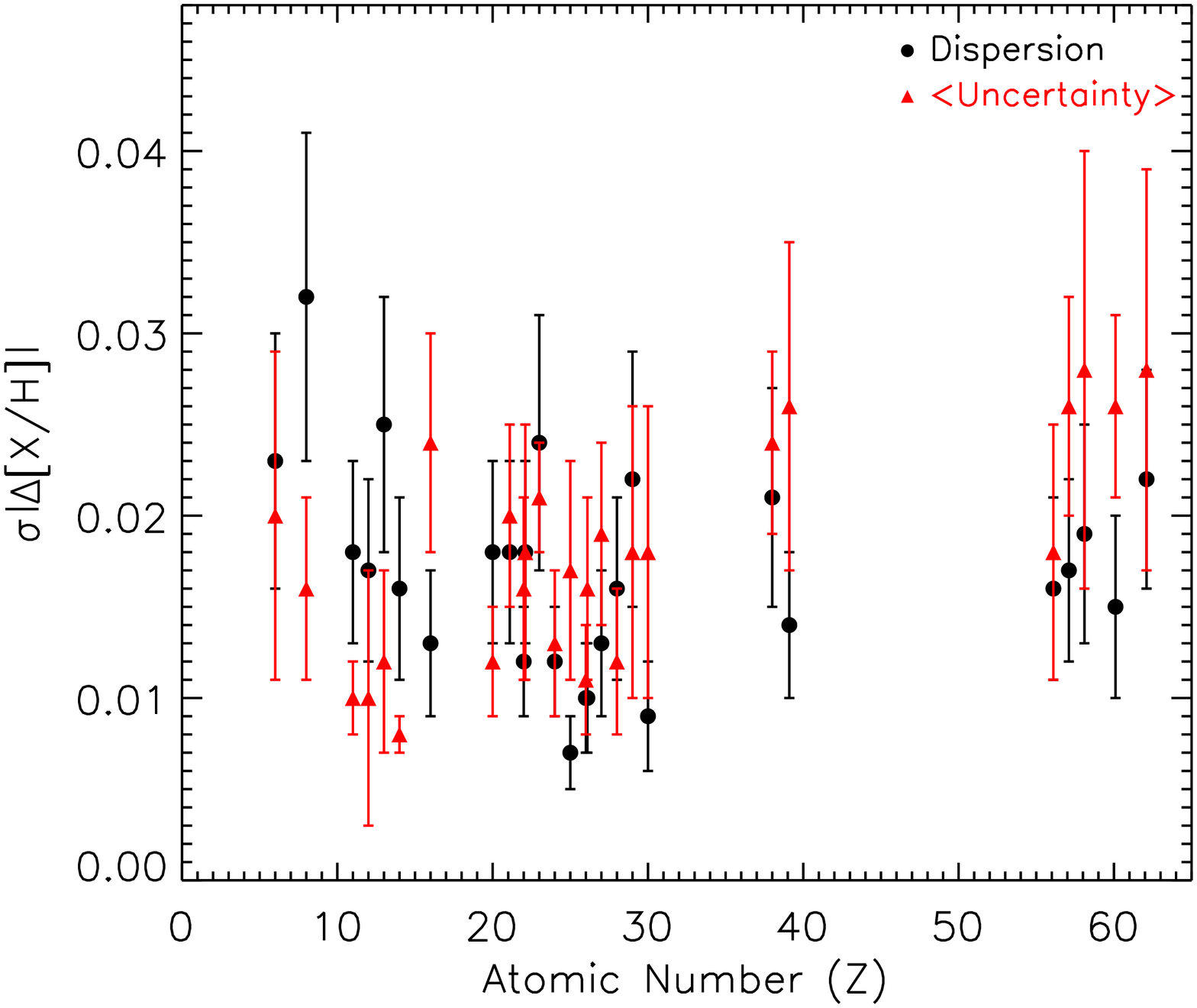}
\caption{The dispersion in the absolute difference of [X/H] abundance ratios between binary components (black circles). The average uncertainty in each element is shown as red triangle.}
\label{fig:disp}
\end{figure}

Binary systems with separation between a few hundreds AU to $\sim$ 1\,pc are believed to form from the same molecular cloud and they are not expected to interact significantly over their lifetimes \citep{sod10,fk14}. Therefore our programme binary systems provide an ideal laboratory to test the chemical homogeneity of these 'mini-clusters'. We note that all of our programme binaries have overall abundance differences between 0.01 to 0.03 dex. The dispersion in relative abundances between binary components are consistent with measurement uncertainties at $\approx$\,0.01 - 0.02 dex on an element-to-element basis (see Figure \ref{fig:disp}), indicating these binaries are chemically homogeneous. The amount of abundance dispersion in our programme binaries is less than that in open clusters (0.03 - 0.05 dex) at similar precision level (e.g., \citealp{liu16,spi18}). This implies the effect of imperfect ISM mixing that, within a star forming region, gas is more homogenous at smaller scale (binary) than at larger scale (the whole cluster). 

In order to examine the chemical anomaly of each pair in detail, we calculated the value of reduced chi-square ($\chi^2_{red}$) of our programme binaries, which is defined as:
\begin{equation}
\chi^2_{red} = \frac{1}{N-1}\sum_{i=1}^{N}(\frac{\Delta[X_i/H]}{\sigma X_i})^2
\end{equation}
where $\Delta$[X$_i$/H] is the abundance difference of the $i^{th}$ element within the pair, $\sigma$X$_i$ is its uncertainty and N is the number of elements analysed in a given pair. The $\chi^2_{red}$ values of our programme binaries are listed in Table \ref{t:stat}. As discussed in \citet{and10} and \citet{nag20}, the $\chi^2_{red}$ value has an uncertainty $\sigma_\chi$ = 2/N. We consider a pair with $\chi^2_{red}$ $>$ 1\,$+$\,5\,$\sigma_\chi$ as chemically anomalous. We found that the three pairs with significant \tcond\ trends also have the largest $\chi^2_{red}$ values (2.716 for HD\,111484A/B, 1.506 for HD\,146368A/B and 1.716 for HIP\,92961/62). As discussed before, the chemical anomaly of these pairs might be due to the effects of planet formation or atomic diffusion.

\begin{figure}
\centering
\includegraphics[width=\columnwidth]{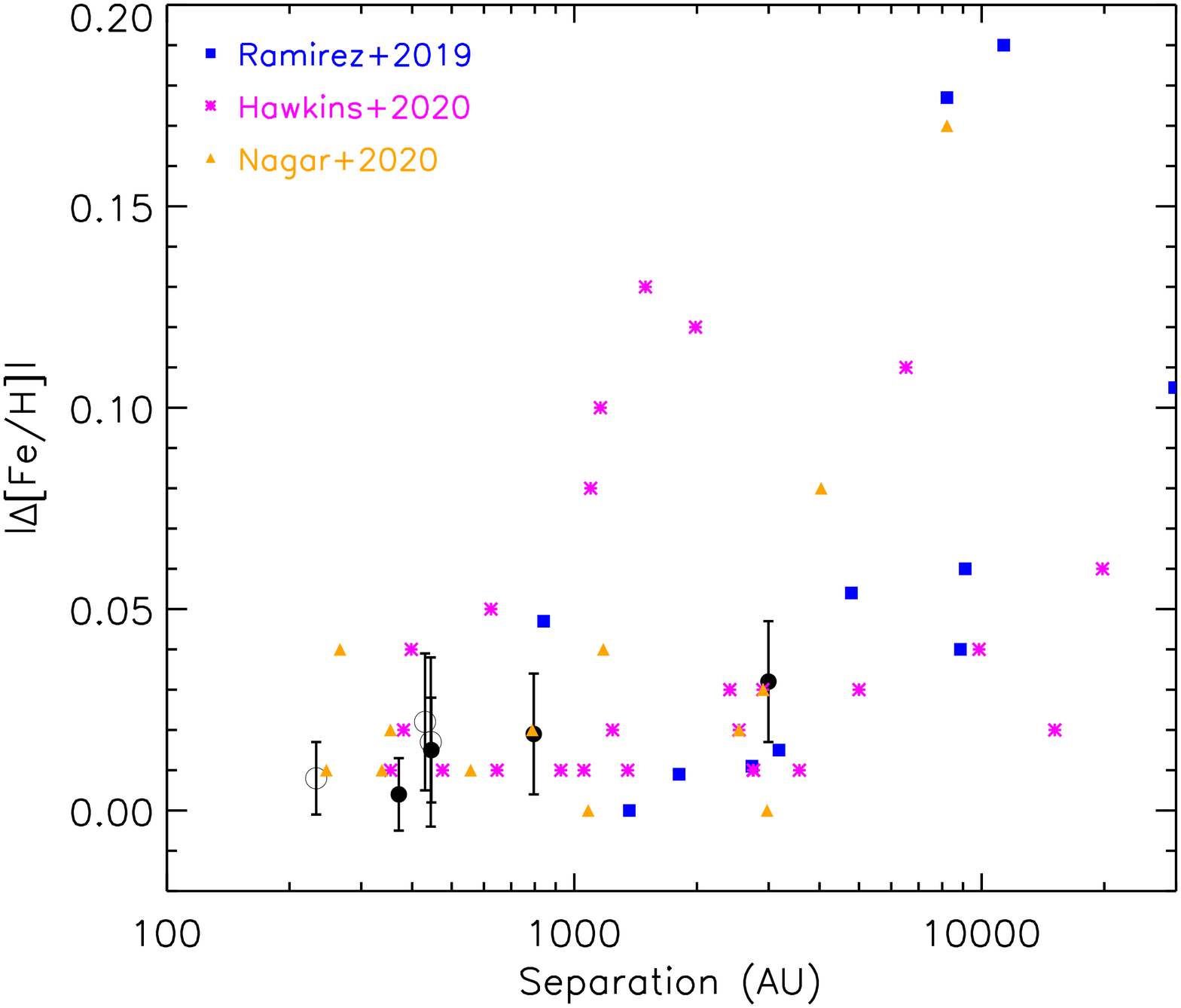}
\caption{Absolute differences in [Fe/H] as a function of binary star projected separation. Black circles represent the results from this work, while the filled and open circles correspond to bound and unbound binary systems. Blue, magenta and orange symbols represent the data from \citet{ram19,haw20} and \citet{nag20}, respectively.} 
\label{fig:dfeh_sep}
\end{figure}

We examined the absolute differences in [Fe/H] within our binary components as a function of projected binary separation in Figure \ref{fig:dfeh_sep}, in a compilation of the data from \citet{ram19,haw20} and \citet{nag20}. Visual inspection appears to indicate a weak correlation between metallicity differences and binary separation: the binaries with wider separation tend to have larger metallicity differences. Indeed, this is consistent with the implication that gas is more homogeneous at smaller scale than at larger scale. Our results thus provide useful constraint on the formation theory of binary star system. However, we are still lacking stars with wider separation ($>$\,0.4\,pc), in order to confirm such a correlation with strong statistical significance.

In addition, we estimated the boundness of our binary systems by checking their locations in log-scale of total velocity difference (log($\Delta$V)) and projected binary separation (log($s$)) following \citet{and17}, where the parallaxes and proper motions of our binaries were adopted from \citet{gai18} as stated. We list the values of $\Delta$V and $s$ in Table \ref{t:stat}. Three binary systems (HD\,106515A/B, HD\,146368A/B and HD\,216122A/B) in our sample are potentially unbound and the remaining four are likely bound. We marked the bound and unbound systems with filled and open circles in Figure \ref{fig:dfeh_sep}. We note that there is no clear distinction in abundance differences between bound and unbound pairs in our sample. 

\begin{figure}
\centering
\includegraphics[width=\columnwidth]{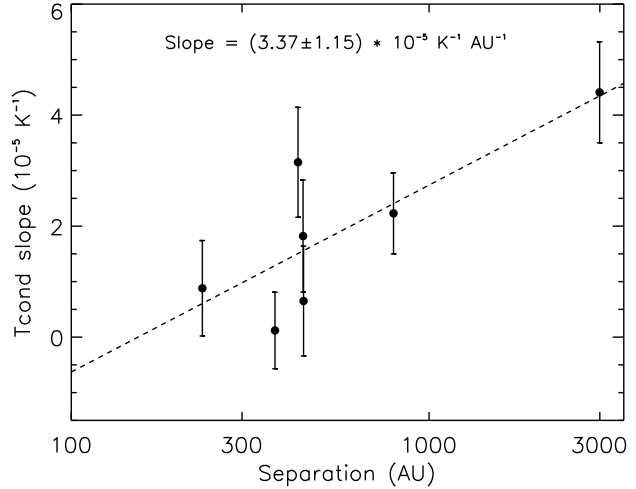}
\caption{Absolute values of the slopes of \tcond\ trends as a function of binary star projected separation for 7 programme binary systems. The dashed line represents the linear least-squares fit to the data.}
\label{fig:test}
\end{figure}

Interestingly, we seem to find that the absolute values of slopes of \tcond\ trends show dependence on the binary separation with 2.9\,$\sigma$ significance, as shown in Figure \ref{fig:test}. It may indicate that the dynamical history of binaries have potential impact on the processes of planet formation. Indeed there was recently the claim that stellar clusters could influence the planet occurrence rates and types of planets \citep{win20}. This effect was also shown in \citet{bru16}, where they found a large excess of hot Jupiters in the dense open cluster M67. Although such a claim is under debate (see \citealt{mus21}), it is possible that the architecture of planetary systems might be affected in binaries to some level as well, since the dynamical interaction and binary separation is relevant. However, we note that the visible correlations may be heavily driven by one pair (HIP\,92961/62) with relatively large binary separation. A larger sample including binaries with large separation ($>$\,1000\,AU and up to 1\,pc) will help us to understand such correlation quantitatively.

\begin{table*}
\caption{Statistical information of the programme binary systems.}
\centering
\label{t:stat}
\begin{tabular}{@{}lrcccrrcc@{}}
\hline\hline
Pair & $<$$\Delta$[X/H]$>$ & $\sigma^{s,a}$ & $<$$\sigma$[X/H]$>$ & $\chi^{2,b}_{red}$ & Slopes$^c$ & [Vol/Ref] & Projected separation & $\Delta$V \\
 & & & & & ($10^{-5}$\,/K) & & (AU) & (km/s) \\
\hline
HD\,106515B$-$A &  0.017 $\pm$ 0.004 & 0.020 & 0.017 & 0.885 &  0.88 $\pm$ 0.86 & $-$0.009 $\pm$ 0.008 & 233 & 2.89 \\
HD\,108575$-$74  &  0.033 $\pm$ 0.004 & 0.021 & 0.021 & 1.239 & $-$0.65 $\pm$ 0.99 & $-$0.001 $\pm$ 0.015 & 446 & 1.14 \\
HD\,111484B$-$A  &  0.017 $\pm$ 0.003 & 0.019 & 0.019 & 2.716 &  2.23 $\pm$ 0.73 & $-$0.018 $\pm$ 0.009 & 795 & 1.35 \\
HD\,133131B$-$A  & $-$0.012 $\pm$ 0.002 & 0.013 & 0.016 & 0.577 &  0.12 $\pm$ 0.69 & 0.003 $\pm$ 0.008 & 371 & 1.46 \\
HD\,146368B$-$A & 0.025 $\pm$ 0.005 & 0.026 & 0.022 & 1.506 & $-$3.15 $\pm$ 0.99 & 0.022 $\pm$ 0.016 & 430 & 2.26 \\
HD\,216122B$-$A  & $-$0.008 $\pm$ 0.004 & 0.022 & 0.024 & 0.373 &  1.82 $\pm$ 1.01 & $-$0.017 $\pm$ 0.006 & 444 & 4.21 \\
HIP\,92962$-$61  &  0.026 $\pm$ 0.005 & 0.027 & 0.021 & 1.716 &  4.41 $\pm$ 0.91 & $-$0.034 $\pm$ 0.019 & 2996 & 0.99 \\
\hline
\end{tabular}
\\
$^a$ Standard deviation of $<$$\Delta$[X/H]$>$. \enspace \enspace \enspace
$^b$ Reduced $\chi^2$ for each pair. \enspace \enspace \enspace
$^c$ Slopes of \tcond\ trends.\\
\end{table*}

\section{Conclusions and future work}

We conducted a high-precision differential analysis of 7 binary star systems from \citet{des04,des06}, using high resolution and high $S/N$ spectra from VLT/UVES and Keck/HIRES. Two pairs (HD\,106515A/B and HD\,133131A/B) host known giant planet(s). We determined the stellar atmospheric parameters of the 'B' component relative to the 'A' component, with very high precision. The differential elemental abundances of up to 30 elements were derived for each pair of binary components in our sample. We managed to attain extremely high precision ($<$\,3.5\%) in the stellar differential abundances.

We found that among the 7 pairs, 4 of them (HD\,106515A/B, HD\,108574/75, HD\,133131A/B and HD\,216122A/B) show subtle abundance differences (0.01 - 0.03 dex) without clear correlations with the condensation temperature (significance level $<$\,2\,$\sigma$), including two pairs hosting known planets (HD\,106515A/B and HD\,133131A/B). Meanwhile another 3 pairs (HD\,111484A/B, HD\,146368A/B and HIP92961/62) show clear \tcond\ trends (significance level $>$\,3\,$\sigma$) with similar degree of abundance differences. 

We do not find any strong evidence of the occurrence of planet versus \tcond\ trend in our sample. This implies that (giant) planet formation does not necessarily imprint chemical signatures and the mass of planet does not necessarily produce the corresponding scale of abundance differences. Although the observed variety in abundance differences with condensation temperature trends might associate with variations of planet formation location (inside or outside H$_2$O/CO ice-line). The results further complicate the current scenario of planet formation and their implications to stellar abundances.

We compared the observed metallicity and abundance offsets with stellar evolutionary models for our binary systems. We found that 4 binaries with surface gravity \logg\ $>$\,0.05 dex (HD\,108574/75, HD\,146368A/B, HD216122A/B and HIP\,92961/62) agree with the models qualitatively. We conclude that the overall abundance offsets between the binary components of these 4 pairs could be due to the effects of atomic diffusion.

In addition, we examined the chemical anomaly in our sample and the correlation between abundance differences and binary separations. We note that our programme binaries have abundance dispersion consistent with typical measurement errors at $\approx$ 0.01 - 0.02 dex, demonstrating that they are chemically homogeneous in general. Although 3 of 7 pairs might be identified as chemical anomaly with the higher $\chi^2_{red}$ values, which could be due to the planet-induced or atomic diffusion effects. The metallicity differences of binary components exhibit weak dependence on the binary separation, which may provide new constraint on the formation theory of binary star systems. We summarise the results of our binary systems in Table \ref{t:sum}.

\begin{table*}
\caption{Summary of our binary systems.}
\label{t:sum}
\begin{tabular}{lccccc}
System & Giant planet(s) & \tcond-abundance correlation & Effects of atomic diffusion & Chemical anomaly & Bound \\
  &  & Sect. 4.1 and 4.2 & Sect. 4.3 & Sect. 4.4 & Fig. \ref{fig:dfeh_sep} \\
\hline
HD\,106515A/B & \checkmark &  &  &  &  \\
HD\,108574/75 &  &  & \checkmark &  & \checkmark \\
HD\,111484A/B & & \checkmark &  & \checkmark & \checkmark \\
HD\,133131A/B & \checkmark &  &  &  & \checkmark \\
HD\,146368A/B &  & \checkmark & \checkmark & \checkmark &  \\
HD\,216122A/B &  & (\checkmark)$^a$ & \checkmark &  &  \\
HIP\,92961/62 &  & \checkmark & \checkmark & \checkmark & (\checkmark)$^a$ \\
\hline
\end{tabular}
\\
$^a$ The correlation or boundness is uncertain.\\ 
\end{table*}

We note the current studies of binary systems are limited. The sample of binary twin stars with high-precision spectroscopic studies are small. Currently only a dozen of binaries have been studied in detail and many of them are lack of complete planet information. This strongly restrict our capability to establish a statistical significant connection between stellar abundance pattern and planet formation. Therefore a comprehensive, high-precision spectroscopic survey of a large number of binaries (preferably $>$ 100) hosting different types of planets (super-Earths, Neptune-like planets, cool and hot giants) will be beneficial to the community. Such a sample will enable us to reveal clearly the planet signatures and constrain the planet formation theory.

Furthermore, stronger observational tests of atomic diffusion, and a potential determination of the ad-hoc turbulent mixing parameter, would come from binaries consisting of a turn-off star and a subgiant star. Such a sample are reasonable to choose from with the recent Gaia Early Data Release 3 \citep{gai21}. We note a large numbers of co-moving binaries with much wider separations have been identified by \citet{elb21}. A high-precision differential analysis of a complete binary sample (with e.g., Gaia G $<$ 10\,mag) in the log($\Delta$V)) - log(s) plane (up to 1\,pc) will enable us to study the effects of atomic diffusion, chemical homogeneity of wide binaries and the dependence on the binary separation, thus providing a well-established reference sample.

\section*{Acknowledgments}
FL and MTM acknowledge the support of the Australian Research Council through Future Fellowship grant FT180100194. MA and DY acknowledge the support of the Australian Research Council through grants FL110100012, FT140100554 and DP120100991. BB thanks the European Research Council (ERC Starting Grant 757448-PAMDORA) for their financial support. BBL is supported by start-up grant of Bairen program from Zhejiang University. YST is grateful to be supported by the NASA Hubble Fellowship grant HST-HF2-51425.001 awarded by the Space Telescope Science Institute. SF acknowledges support by the grant "The New Milky Way" from the Knut and Alice Wallenberg Foundation.
The authors thank the ANU Time Allocation Committee for awarding observation time to this project. The authors wish to acknowledge the very significant cultural role and reverence that the summit of Mauna Kea has always had within the indigenous Hawaiian community. Finally, the authors are grateful for the valuable comments from the referee that helped improve the manuscript.

\section*{Data availability}
The spectral data underlying this article are available in Keck Observatory Archive at \url{https://koa.ipac.caltech.edu/cgi-bin/KOA/nph-KOAlogin} and ESO Science Archive Facility at \url{ http://archive.eso.org/eso/eso_archive_main.html}, and can be accessed with Keck Program ID: Z007Hr (Semester: 2016A, PI: Asplund) and ESO Programme ID: 101.D-0921(A) (PI: Liu), respectively.\\
The rest of data underlying this article are available in the article and in its online supplementary material.

\section*{SUPPLEMENTARY MATERIAL}

The following material is available online for this article: \\
Table A1. Atomic line data and the EW measurements adopted for our analysis.

\label{lastpage}

\end{CJK*}
\end{document}